\documentclass[a4paper,oneside,11pt]{article}
\usepackage{amsmath}

\usepackage{amsmath,amstext,amsfonts,amsbsy,amssymb,amscd,bbm,epsfig,lscape}

\usepackage[titletoc,title]{appendix}
\usepackage{hyperref}

\usepackage{times}
\usepackage{float}

\usepackage{subfig}
\usepackage{graphicx}

\usepackage{cprform}

\usepackage{authblk}

\usepackage{cite}

\usepackage{enumitem}

\newcommand{\Dslash}{\relax{\kern+.25em / \kern-.70em D}}

\newcommand{\Real}{\relax{\mathsf{\Gamma\kern-.35em R}}}
\newcommand{\Int}{\relax{\mathsf{Z\kern-.40em Z}}}

\newcommand{\be}{\begin{equation}}
\newcommand{\ee}{\end{equation}}
\newcommand{\bea}{\begin{eqnarray}}
\newcommand{\eea}{\end{eqnarray}}

\newcommand{\obar}[1]{\kern3pt\overline{\kern-2pt #1\kern-0pt}\kern1pt}

\newcommand{\corrbar}[1]{\kern3pt\overline{\kern-2pt #1\kern-0pt}\kern1pt}

\newcommand{\oVApAVren}[1]{\kern3pt\overline{\kern-2pt #1\kern-0pt}\kern1pt_{\rm\scriptscriptstyle VA+AV;s}}

\newcommand{\zbar}{\kern3pt\overline{\kern-2pt Z\kern-0pt}\kern1pt}

\newcommand{\zbarVApAV}[1]{\kern3pt\overline{\kern-2pt Z\kern-0pt}\kern1pt_{\rm\scriptscriptstyle VA+AV #1}}

\begin{document}


\begin{titlepage}


\vspace*{0truemm}


\centerline{{\large \bf  Mass of the $\mathbf{b}$-quark  and $\mathbf{B}$-decay constants
from  $\mathbf{N_f=2+1+1}$ twisted-mass Lattice QCD }} 

\vskip 5 true mm
\centerline{\bigrm A.~Bussone$^{(a)}$, N.~Carrasco$^{(b)}$, P.~Dimopoulos$^{(c,\,d)}$, R.~Frezzotti$^{(d,\,e)}$, P.~Lami$^{(f)}$, 
}
\vspace*{0.2cm}
\centerline{\bigrm   V.~Lubicz$^{(b,\,f)}$, E. Picca$^{(f)}$, L. Riggio$^{(b)}$, G.C.~Rossi$^{(c,\,d,\,e)}$, S.~Simula$^{(b)}$, 
C.~Tarantino$^{(f,\,b)}$} 

\vspace*{0.2cm}
\centerline{\bigrm  (for the ETM Collaboration) }

\vskip 1 true mm
\vskip 0 true mm

\centerline{\it $^{(a)}$ CP3-Origins \& Danish IAS, University of Southern Denmark}
\centerline{\it Campusvej 55, 5230 Odense M, Denmark}
\vskip 2 true mm
\centerline{\it $^{(b)}$ INFN, Sezione di Roma Tre}
\centerline{\it Via della Vasca Navale 84, I-00146 Rome, Italy}
\vskip 2 true mm
\centerline{\it $^{(c)}$ Centro Fermi - Museo Storico della Fisica e Centro Studi e Ricerche Enrico Fermi}
\centerline{\it Compendio del Viminale, Piazza del Viminiale 1, I-00184, Rome, Italy}
\vskip 2 true mm
\centerline{\it $^{(d)}$ Dipartimento di Fisica, Universit\`a di Roma ``Tor Vergata''}
\centerline{\it Via della Ricerca Scientifica 1, I-00133 Rome, Italy}
\vskip 2 true mm
\centerline{\it $^{(e)}$ INFN, Sezione di ``Tor Vergata"}
\centerline{\it Via della Ricerca Scientifica 1, I-00133 Rome, Italy}
\vskip 2 true mm
\centerline{\it $^{(f)}$ Dipartimento di Fisica, Universit\`a  Roma Tre}
\centerline{\it Via della Vasca Navale 84, I-00146 Rome, Italy}
\vskip 2 true mm

\vskip 20 true mm


\thicktablerule
\vskip 3 true mm
\noindent{\tenbf Abstract}
\vskip 1 true mm
\noindent
{\tenrm  
We present precise lattice computations for the $b$-quark mass, the quark mass ratios $m_b/m_c$ and $m_b/m_s$ as well as the 
leptonic $B$-decay constants. 
We employ  gauge configurations with four dynamical quark flavors, up/down, strange and charm,  
at three values of the lattice spacing ($a \sim 0.06 - 0.09$ fm) and for pion masses as low as 210 MeV. 
Interpolation in the heavy quark mass to the bottom quark point is performed using ratios of physical quantities computed
at nearby quark masses exploiting the fact that these ratios are exactly known in the static quark mass limit.
Our results are also extrapolated to the physical pion mass and to the continuum limit and  read: 
$m_b(\overline{{\rm MS}}, m_b)  = 4.26(10)$ GeV, $m_b / m_c  = 4.42(8)$, 
$m_b / m_s  = 51.4(1.4)$, $f_{Bs} = 229(5)$ MeV, $f_B = 193(6)$ MeV, $f_{Bs}/f_B = 1.184(25)$ and 
$(f_{Bs}/f_B) / (f_K/f_{\pi}) = 0.997(17)$.  
}
\vskip 3 true mm
\thicktablerule
\eject
\end{titlepage}

\section{Introduction}
\label{sec:intro}
Lattice QCD simulations constitute the current dominant theoretical framework for high-precision $B$-physics computations which 
are  necessary, in combination with experimental results,  to obtain precious information in quark sector phenomenology. 
In fact increasingly improved computations of matrix elements (decay constants, form factors and mixing parameters)  
are of high importance to carry out challenging tests of the  
Cabibbo-Kobayashi-Maskawa (CKM) paradigm, an effort also stimulated by the ambitious prospect  
of discovering footprints of New Physics effects.
Moreover lattice methods are optimal to determine  the quark masses  by confronting experimental quantities from spectroscopy with 
their theoretical counterparts computed from first principles via lattice QCD simulations.  

We should stress that although direct lattice simulations are not yet possible at the physical value of the $b$-quark mass due to computing power limitations, 
the combined use of effective theories and improved lattice techinques has progressively led to results that are characterised  by much 
reduced and reliable systematic uncertainties.    

In the present paper we have carried out a non-perturbative determination of the $b$-quark mass as well as its 
ratios to the charm and the strange quark mass. The latter turn out to be very accurate because 
the renormalisation scheme dependence is absent 
and the systematics  related to the lattice scale  
determination are suppressed. 
We observe that a precise $b$-quark mass evaluation is important for reducing the uncertainty in the study of 
Higgs decays to $b \bar{b}$~\cite{Djouadi:2005gi} and possibly unveal non-SM features of the $H$-$b$-$\bar{b}$ coupling. 

In this paper we have also computed the pseudoscalar $B$-decay constants $f_{Bs}$ and $f_B$ as well as their (SU(3)-breaking) ratio, $f_{Bs}/f_B$. 
Currently there is high experimental interest by LHCb and  $B$-factories in 
the processes $B_{(s)} \rightarrow \mu^{+}\mu^{-}$~\cite{CMS:2014xfa} and 
$B \rightarrow \tau \nu$~\cite{Adachi:2012mm,Lees:2012ju } for the  full 
description of which the knowledge of the aforementioned decay constants is indispensable.
The importance of $B$-decays is not limited only to their crucial contribution for improving the 
accuracy of the unitarity triangle determination; 
in fact $B$-decays in channels that are loop suppressed in SM  are some of the 
first-class candidates for revealing features of beyond the Standard Model (SM) dynamics.    

In our lattice computation we have used $N_f=2+1+1$ dynamical quark gauge configurations generated by 
ETM Collaboration~\cite{Baron:2010bv,Baron:2010th} 
at three values of the lattice spacing. Our results are extrapolated to the continuum limit.   
For the determination  of the $B$-physics observables we have employed the ETMC {\it ratio method} 
that has already been applied within  the $N_f=2$ lattice 
simulations framework~\cite{Blossier:2009hg, Dimopoulos:2011gx, Carrasco:2013zta}.     
In particular in the present paper we have brought about improvements of the ratio method implementation thanks to which 
it is possible to gain better control 
on various sources of systematic uncertainty.    

The plan of the paper is the following. We describe our computational setup in Section~\ref{sec:compu-setup}. 
In Section~\ref{sec:analy_results} we present 
an improved implementation of the ratio method in the cases of the determination of the $b$-quark mass, its 
ratios to the charm and strange quark masses,  
and the pseudoscalar $B$-decay constants. We also give a detailed error budget for each one of the 
observables studied in the present work. 
Finally in  Section~\ref{sec:concl} we compare our results with the ones provided by other 
lattice collaborations. For the interested reader, recent reviews on 
$B$-physics lattice computational methods, techniques and collection of results 
are given in Refs.~\cite{Aoki:2013ldr, Bouchard:2015pda, Carloslat15, Rosner:2015wva}. Recent non-lattice results 
can be found {\it e.g.} in Refs.~\cite{Chetyrkin:2010ic, Narison:2012mz, Lucha:2013gta}.

\section{Computational details}
\label{sec:compu-setup}

\subsection{Lattice action setup}
In our computation we employ Iwasaki glue~\cite{Iwasaki:1985we, Iwasaki:1996sn} and  a mixed lattice fermionic action setup. 
The sea quark action for the light mass-degenerate sea quark doublet, $S_{\ell}$, and the action for the strange and 
charm quark doublet, $S_h$  (~\cite{FrezzoRoss1, Frezzotti:2003xj}) 
read, respectively,
\begin{align}
S_{\ell}^{sea} &=a^4\sum_{x}\bar{\psi}_{\ell}(x)\left\{ \dfrac{1}{2}\gamma_{\mu}\left(\nabla_{\mu}+
\nabla_{\mu}^{*}\right)-i\gamma_{5}\tau^{3}\left[M_{\textrm{cr}}-\dfrac{a}{2}\sum_{\mu}\nabla_{\mu}^{*}
\nabla_{\mu}\right]+\mu_{\ell}\right\} \psi_{\ell}(x), \label{eq:tmlight} \\ 
S_{h}^{sea} &=a^4\sum_{x} \bar{\psi}_{h}(x)\left\{ \dfrac{1}{2}\gamma_{\mu}\left(\nabla_{\mu}+\nabla_{\mu}^{*}\right)- 
i\gamma_{5}\tau^{1}\left[M_{\textrm{cr}}-
\dfrac{a}{2}\sum_{\mu}\nabla_{\mu}^{*}\nabla_{\mu}\right]+{\mu}_{\sigma}+{\mu}_{\delta}{\tau}^{3} \right\}  {\psi}_{h}(x),
\label{eq:tmheavy} 
\end{align}
where $\nabla_{\mu}$ and $\nabla_{\mu}^{*}$ represent the nearest neighbour forward and backward covariant 
derivatives and it is intended that the untwisted mass has been tuned to its critical value, $M_{cr}$. 
In Eq.~(\ref{eq:tmlight}) we have defined the quark doublet $\psi_{\ell} = (\psi_u~ \psi_d)^{T}$ while  $\mu_{\ell}$ 
denotes the light (sea) twisted quark mass.    
In Eq.~(\ref{eq:tmheavy}) $\psi_h=(\psi_s, \psi_c)^T$ denotes the strange-charm fermion doublet while 
$\mu_{\sigma}$ and $\mu_{\delta}$ are the bare twisted mass parameters 
from which the renormalized (sea) strange and charm  
masses can be derived. Pauli matrices in Eqs~(\ref{eq:tmlight}) and~(\ref{eq:tmheavy}) act in flavor space. 
For more details on the twisted mass setup we refer the reader to Refs.~\cite{FrezzoRoss1, Frezzotti:2003xj, 
Baron:2010bv, Baron:2010th, Boucaud:2007uk, Boucaud:2008xu, Baron:2009wt}.
 
In the valence sector we employ the Osterwalder-Seiler (OS) action~\cite{Osterwalder:1977pc} 
which is written as the sum of  {\it individual} quark flavor contributions 
\begin{equation}\label{eq:OS}
S_{q}^{val, \textrm{OS}}=a^4\sum_{x}\sum_{f}\bar{q}_{f}\left\{ \dfrac{1}{2}\gamma_{\mu}\left(\nabla_{\mu}+\nabla_{\mu}^{*}\right)-i\gamma_{5}r_{f}
\left[M_{\textrm{cr}}-\dfrac{a}{2}\sum_{\mu}\nabla_{\mu}^{*}\nabla_{\mu}\right]+\mu_{f}\right\} q_{f}(x) \, , 
\end{equation}
where the label $f$ runs over the different valence flavors light, strange, charm or heavier and $r_f = \pm 1$.  
Valence and sea quark masses are matched to each other and fixed in terms of meson masses in order to ensure unitarity in the 
continuum limit. Lattice artifacts in physical observables are just O$(a^2)$~\cite{FrezzoRoss1, Frezzotti:2004wz}.

\subsection{Simulation parameters and correlation functions}

We have used the $N_f=2+1+1$  gauge ensembles generated 
by the ETM Collaboration~\cite{Baron:2010bv,Baron:2010th}. 
A  summary of the most important details of our simulations is given in Table~\ref{tab:runs}.         
\begin{table}[!h] 
\begin{center}
\scalebox{0.90}{
\begin{tabular}{||c|c|c|c|c|c||}
\hline
 $\beta$ & $V / a^4$ &$a\mu_{sea}=a\mu_\ell$ & $N_{cfg}$& $a\mu_s$& $a\mu_c - a\mu_h$ \\
\hline
 $1.90$ & $32^{3}\times 64$ &$0.0030$ &$150$ &  $0.0180,$& $0.21256, 0.25000,$ \\
        & & $0.0040$ &  $150$ &$0.0220,$ & 0.29404, 0.34583, \\
        & & $0.0050$ &   $150$ &$ 0.0260$ &  0.40675, 0.47840, \\
        & &          &         &          &  0.56267, 0.66178, \\
        & &          &         &          &  0.77836,  0.91546 \\
\cline{1-4}
 $1.90$ & $24^{3}\times 48 $ & $0.0040$ & $150$ & & \\
        & & $0.0060$ &   $150$ &&  \\
        & & $0.0080$ &   $150$ & &  \\
        & & $0.0100$ &   $150$ & &  \\
\hline
 $1.95$ & $32^{3}\times 64$ &$0.0025$ & $150$& $0.0155,$& $0.18705, 0.22000,$ \\
        & & $0.0035$ &  $150$ &$  0.0190,$ &0.25875, 0.30433,\\
        & & $0.0055$ &  $150$ &$ 0.0225$    &0.35794, 0.42099, \\
        & & $0.0075$ &  $150$  &             &0.49515, 0.58237   \\
        & &          &        &             &0.68495,  0.80561 \\
\cline{1-4}
 $1.95$ & $24^{3}\times 48 $ & $0.0085$  & $150$ & & \\
\hline
 $2.10$ & $48^{3}\times 96$ &$0.0015$ & $90$& $0.0123,$& $0.14454, 0.17000, $ \\
        & & $0.0020$ &   $90$ &$0.0150,$ &0.19995, 0.23517, \\
        & & $0.0030$ &  $90$ & $  0.0177$  &0.27659, 0.32531,  \\
        & &          &       &             &0.38262, 0.45001, \\
        & &          &       &             &0.52928,  0.62252 \\
 \hline
\end{tabular}
}
\caption{\label{tab:runs} Summary of simulation details. Gauge couplings $\beta$ = 1.90, 1.95 and 2.10 correspond 
to lattice spacings $a \simeq$ 0.089, 0.082 and 
0.062 fm, respectively. We denote with $a\mu_{\ell}$, $a\mu_{s}$ and $a\mu_{c}-a\mu_{h}$, the light, 
strange-like, charm-like and  heavier bare quark masses, respectively, entering in the valence sector computations. 
$N_{cfg}$ stands for the number of gauge configurations used in the analysis. } 
\end{center}\label{Tab1}
\end{table} 

Simulation data have been taken at three values of the lattice spacing, 
namely $a=0.0885(36)$, $0.0815(30)$ and $0.0619(18)$~fm, 
corresponding to $\beta=1.90$, $1.95$ and $2.10$, respectively (see Ref.~\cite{Carrasco:2014cwa}). 
In our simulation the light valence and sea quark masses are set equal, leading to pion masses in the range between 210 and 
450~MeV. Strange and charm sea quark masses are chosen close to their physical value and fixed from $M_K$ and $M_{Ds}$ inputs 
(see Ref.~\cite{Carrasco:2014cwa}). 
To allow for a smooth interpolation to the physical values of the valence strange and charm mass as well as for heavier quark masses, 
we have inverted the heavy valence Dirac matrix for three  values of the strange-like quark mass, $\mu_{s}$, 
and a number  of charm-like  and heavier quark mass, $a\mu_{c}-a\mu_{h}$. 

We have fixed the lattice scale using $f_{\pi}$. The $u/d$, strange and charm quark masses
have been determined comparing lattice data with the experimental values of the pion, $K$ and $D_{(s)}$ 
meson mass, respectively. Further details can be found in Ref.~\cite{Carrasco:2014cwa}. 
The use of the mixed action of twisted mass and OS quarks offers the advantage 
that the masses of light quarks in the sea and of all
types of quarks in the valence are multiplicatively renormalised via the renormalisation constant (RC) $Z_m = 1/Z_P$. 
The latter  is computed nonperturbatively using the RI -MOM scheme 
(for the RC determination see Appendix A of Ref.~\cite{Carrasco:2014cwa}). 
Moreover, exact chiral lattice Ward-Takahashi identities imply that at maximal twisted 
angle no normalisation constant is needed in 
the computation of decay constants~\cite{Frezzotti:2000nk,FrezzoRoss1}.

\begin{figure}%
    \centering
    \subfloat[]{{\includegraphics[width=7.5cm]{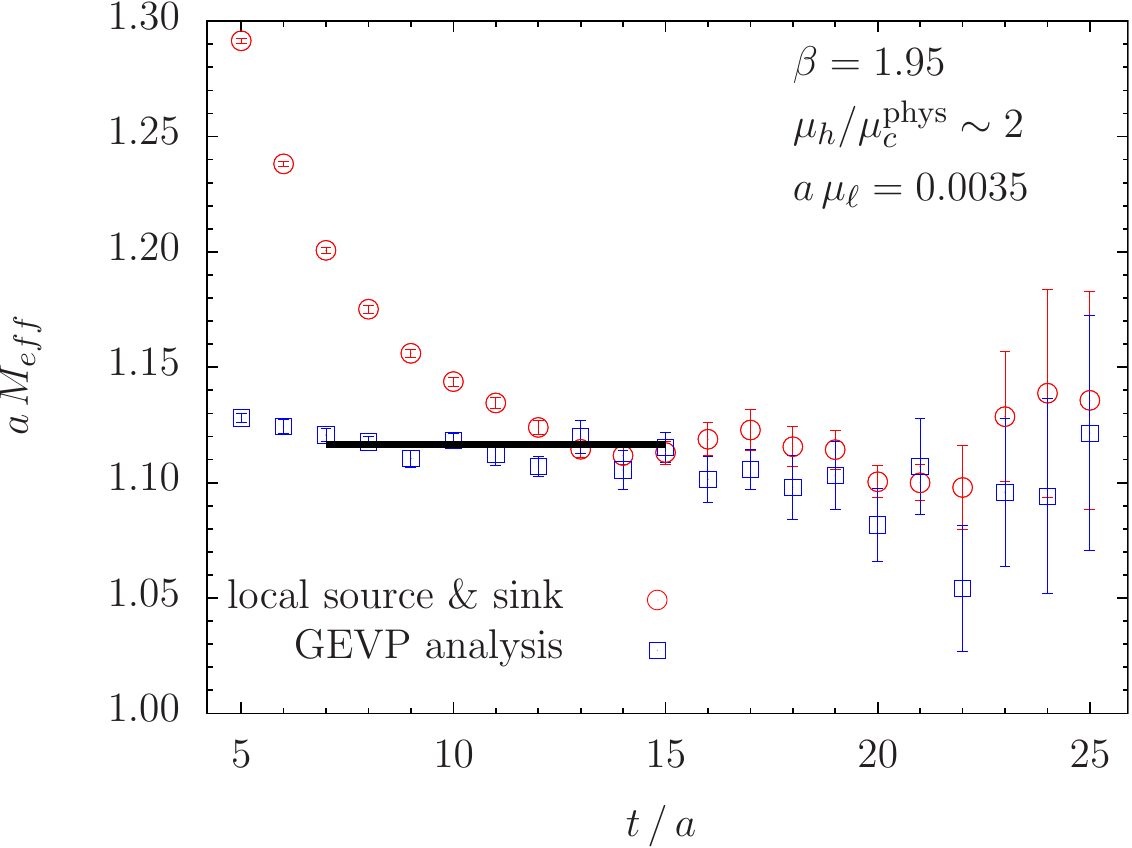} }}%
    \qquad
    \subfloat[]{{\includegraphics[width=7.5cm]{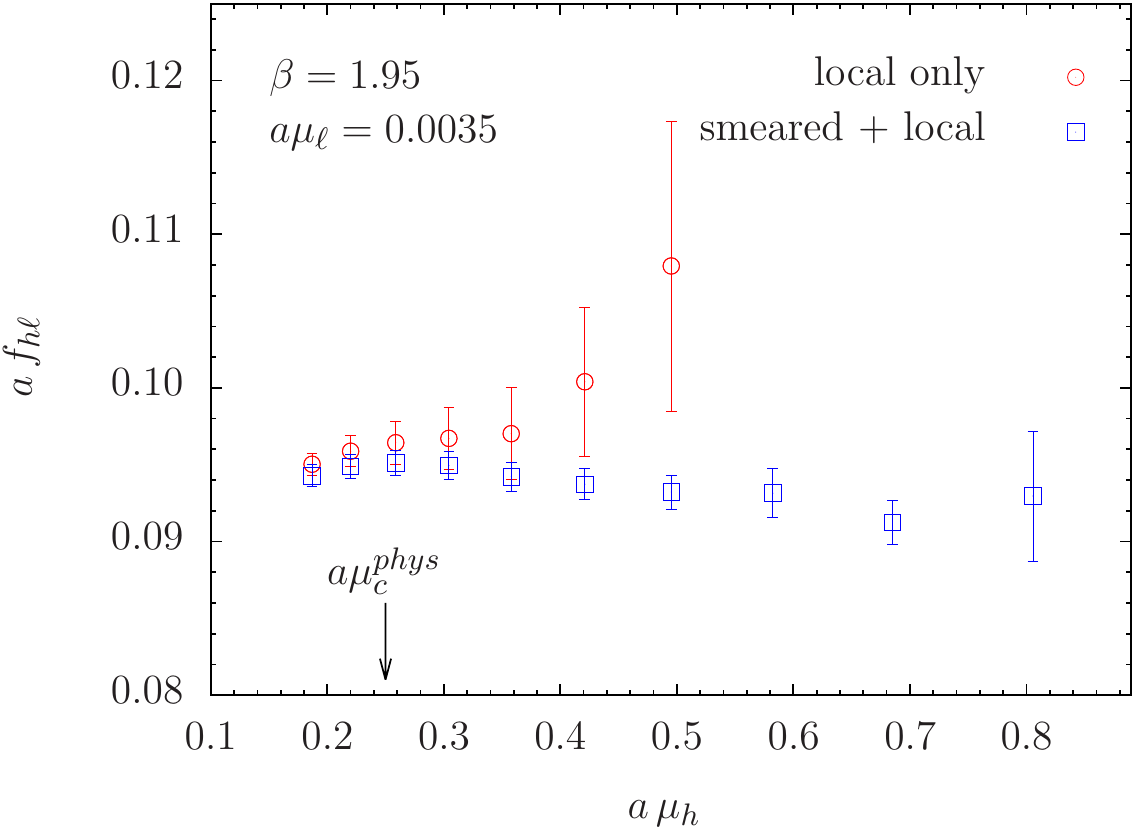} }}%
    \caption{(a) Effective mass of the pseudoscalar 2-point correlator obtained using either a local source and sink (red circles)
or the GEVP method (blue squares) applied to a matrix of local-local, smeared-local, local-smeared and smeared-smeared correlators  vs. the Euclidean time separation in lattice units.
Here $\beta =1.95$,    $a\mu_{\ell} = 0.0035$ on lattice volume $V/a^4 = 32^3 \times 64$. 
The heavy-quark mass is around two times the physical charm quark mass, $\mu_c^{phys}$.
(b) The decay constant of heavy-light mesons computed using only local interpolating fields (red circles) or including Gaussian
smeared sources (blue squares) are plotted vs. the heavy-quark mass, ranging from $\mu_c^{phys}$ up to $\sim 3\,\mu_c^{phys}$. }
    \label{fig:Meff_fhl}
\end{figure}

In two-fermion correlation functions valence light and strange-like quark propagators 
have been calculated with the ``one-end" trick stochastic method~\cite{Foster:1998vw, McNeile:2006bz} 
by employing spatial stochastic sources at a randomly chosen time-slice.
However for propagators of the charm  or heavier quark, in order to get suppressed contribution 
of the excited states in the correlation functions, we have employed Gaussian smeared
interpolating quark fields~\cite{Gusken:1989qx}. 
For the values of the smearing parameters we set $k_{G}=4$ and $N_{G}=30$. In addition, we apply APE-smearing 
to the gauge links~\cite{Albanese:1987ds} in the interpolating fields with parameters $\alpha_{APE}=0.5$ and $N_{APE}=20$.
Smearing leads to improved projection onto the lowest energy eigenstate at small Euclidean time separations. 
We have implemented smeared fields in both source and sink. 
We have thus evaluated two-point heavy-light correlation 
functions made up by the four possible combinations 
of local/smeared source/sink. We can thus  employ  the GEVP method~\cite{Blossier:2009kd} to 
compute the ground state pseudoscalar masses.
For the pseudoscalar decay constant calculation we evaluate two point correlation functions with pseudoscalar interpolating operators 
$P(x) = \overline{q}_1(x) \gamma_5 q_2(x)$. The typical form of the correlation function and its 
asymptotic behaviour on periodic lattices read: 
\begin{equation}
C_{PP}(t) = (1/L^3)\sum_{\vec{x}} \langle P(\vec{x},t) P^{\dagger}(\vec{0}, 0) \rangle 
           \stackrel{t \gg 0, ~(T-t) \gg 0}{\xrightarrow{\hspace*{1.5cm}}}
          \dfrac{\xi_{PP}}{2M_{ps}} \left(e^{-M_{ps} t} + e^{-M_{ps}(T-t)} \right)
\end{equation}
We set opposite Wilson parameters, $r_f$, for the two valence quarks that form the pseudoscalar meson. This choice 
guarantees that the cutoff effects on the pseudoscalar mass are
$O(a^2 \mu_q)$~\cite{FrezzoRoss1, Frezzotti:2005gi, Dimopoulos:2009qv}.
We consider two cases, using smeared source only and
source and sink both smeared, for which $\xi_{PP}$ is given by
$\xi_{PP} = \langle 0| P^{L}|ps \rangle \langle ps |P^{S}|0 \rangle$ in the first case and
$\xi_{PP} = \langle 0 | P^{S}|ps \rangle\langle ps |P^{S}|0\rangle$ in the second one,
where $L$ and $S$ indicate local and smeared operators, respectively.
From the combination of the two kinds of correlators it is easy to get the matrix element of the local operator, namely, 
$g_{ps} = \langle 0 | P^{L} | ps \rangle $ which, via PCAC, allows  for the computation of the pseudoscalar decay constant:
\begin{equation} \label{eq:fps}
f_{ps} = (\mu_{1} + \mu_{2}) \dfrac{g_{ps}}{M_{ps} \sinh M_{ps}}, 
\end{equation}
where $\mu_{1, 2}$  are the masses of the valence quarks entering the pseudoscalar  meson mass $M_{ps}$.
In Eq.~(\ref{eq:fps}) the use of $\sinh M_{ps}$ rather than $M_{ps}$ 
turns out  to be advantageous for getting reduced discretisation errors.
In Fig.~\ref{fig:Meff_fhl} we illustrate the beneficial effect of smearing in determining the ground state signal 
at early time distance and making 
possible the decay constant evaluation for values of the heavy quark mass for which it fails if local interpolating fields 
only are used.     
In the figure we show the results at $\beta=1.95$ and $a\mu_{sea} = a\mu_{\ell}=0.0035$  
obtained for the heavy-light effective mass versus the Euclidean time separation (left panel) and the decay constant 
versus the heavy quark mass (right panel)
using either only local or appropriate combinations of local and smeared interpolating fields. 

For each of the $\beta=1.90$ and 1.95 gauge ensembles ETM Collaboration has produced 
around 5000 thermalised trajectories; 
for the ensembles at $\beta=2.10$ corresponding to the two heaviest light quark masses ({\it i.e.} $a\mu_{sea}=0.0020, 0.0030$) 
4000 trajectories have been generated while for the case of the lightest sea quark mass ($a\mu_{sea}=0.0015$) the total number 
of the generated trajectories is about 2100. All trajectories have integration length $\tau = 1$.  
In each ensemble gauge configurations are saved on one every two trajectories. 
For each hadronic observable the autocorrelation has been studied either by computing directly 
the $\tau_{int}$  or employing the blocking method to estimate the final error; blocking of   
about 25 measurements is the typical case for our ensembles in order to safely estimate the final statistical error 
for $M_{ps}$ and $f_{ps}$ in the light sector. 
Both methods lead to comparable estimates  for the final statistical error. 
Typical values for $\tau_{int} \in [1, 3]$ for $M_{ps}$ and $f_{ps}$ depending on the gauge ensemble. 
In our analysis we have used a number of measurements, indicated by $N_{cfg}$ in Table~\ref{tab:runs}, for $M_{ps}$ and $f_{ps}$ 
performed on gauge configurations each of which is separated
by about 20 gauge configurations (or equivalently separated by about 40 trajectories with $\tau=1$) of the 
original Monte Carlo 
history. Based on the above findings we are confident that this choice ensures that autocorrelation is highly suppressed. 
Moreover we perform our final analysis by applying the blocking method; 
we consider blocks of 10 measurements for $\beta=1.90$ and $\beta=1.95$ 
and 6 measurements for $\beta=2.10$.
 
Statistical errors on pseudoscalar meson masses and pseudoscalar
decay constants have been estimated with the jackknife procedure. 
Autocorrelation is taken into account using the blocking method. 
Fit cross correlations are kept under control by generating
1000 bootstrap samples for each gauge configuration ensemble. Notice also that the RC computation has
been performed on separate ({\it i.e.} totally uncorrelated to the $N_f=2+1+1$ sets) $N_f = 4$ gauge configuration 
ensembles (for details see Appendix A of Ref.~\cite{Carrasco:2014cwa}). 
Moreover, from the comparison of results obtained at the same lattice spacing ($\beta=1.90$) 
and light quark mass ($\mu_{{\rm sea}} = 0.0040$) 
but on different lattice volumes ($24^3 \times 48$ and $32^3 \times 64$) we notice no significant 
finite volume effects on the values of all observables relevant for this study. 
Note that finite size effects are expected to be maximal correspondingly to the $L=24$, $\mu_{\ell}=0.0040$ ensemble 
as it has the smallest value of $(M_{ps}L)$ among those we have considered (see Ref.~\cite{Carrasco:2014cwa}).

\section{Analysis and results}
\label{sec:analy_results}

For the determination of the $B$-physics quantities we have used the {\it ratio method}
already applied in the $N_f=2$ framework~\cite{Blossier:2009hg, Dimopoulos:2011gx, Carrasco:2013zta}.   
The main idea can be summarized in three steps. The first one is the calculation of the values
of the observables of interest at heavy quark masses around the charm scale, for which
relativistic simulations are reliable (i.e. they produce results with  well controlled
discretisation errors). The second step consists in evaluating appropriate ratios of the
observables at increasing values of the heavy quark mass up to a scale of 2-3 times
the charm quark mass (i.e. around 3 GeV). The key point is that the static limit of the measured ratios
is exactly known from HQET arguments. The final step of the computation
consists in smoothly interpolating data from the charm region to the infinite mass point
and extracting their values at the $b$-mass. 

The great computational advantage of this method is that one is able to make $B$-physics computations
using the same relativistic action setup with which the lighter quark computations are
performed. Moreover an extra simulation at the static point limit is not necessary, 
while the relevant exact information about it is incorporated in
the construction of the ratios of observables.

It should be stressed that the use of ratios
of observables drastically reduces the discretisation errors and at the same time  leads to a great suppression of 
the uncertainties that come from the QCD matching to HQET. Furthermore the impact of possible (residual) effects 
of both types of systematic uncertainty  on the final results can be 
controlled by employing appropriate variants of the ratio definition (see below).

First preliminary analyses for the decay constants and the $b$-quark mass with $N_f=2+1+1$ gauge ensembles have been presented 
in Refs.~\cite{Carrasco:2013naa, Bussone:2014cha}, respectively.

In the following sections we present our $B$-physics analysis where we have made use of improved variants of the 
ratio method that allow for better control over three main sources of systematic uncertainty, namely, those due to  
discretisation, lattice scale determination and the fitting procedure related to the $b$-point interpolation.

\subsection{Bottom quark mass and bottom to charm/strange quark mass ratios}
\begin{figure}%
    \centering
    \subfloat[]{{\includegraphics[width=7.5cm]{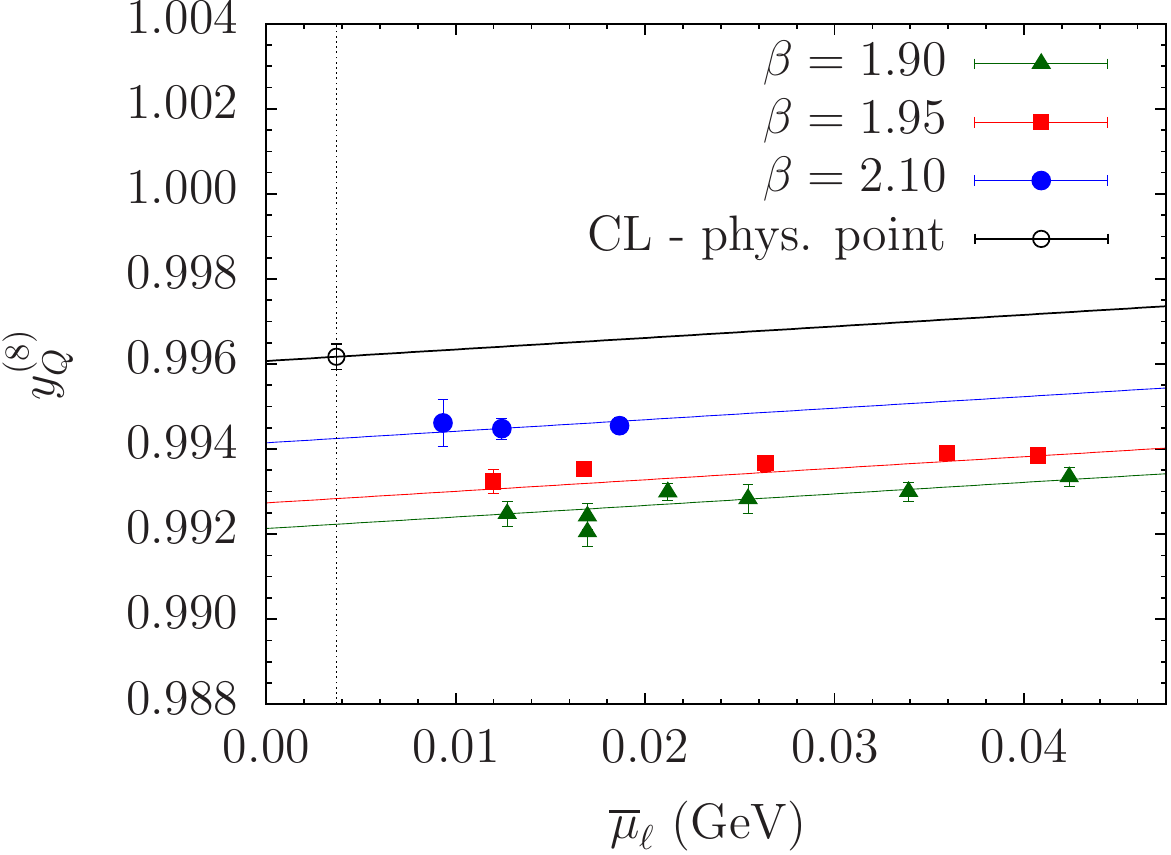} }}%
    \qquad
    \subfloat[]{{\includegraphics[width=7.5cm]{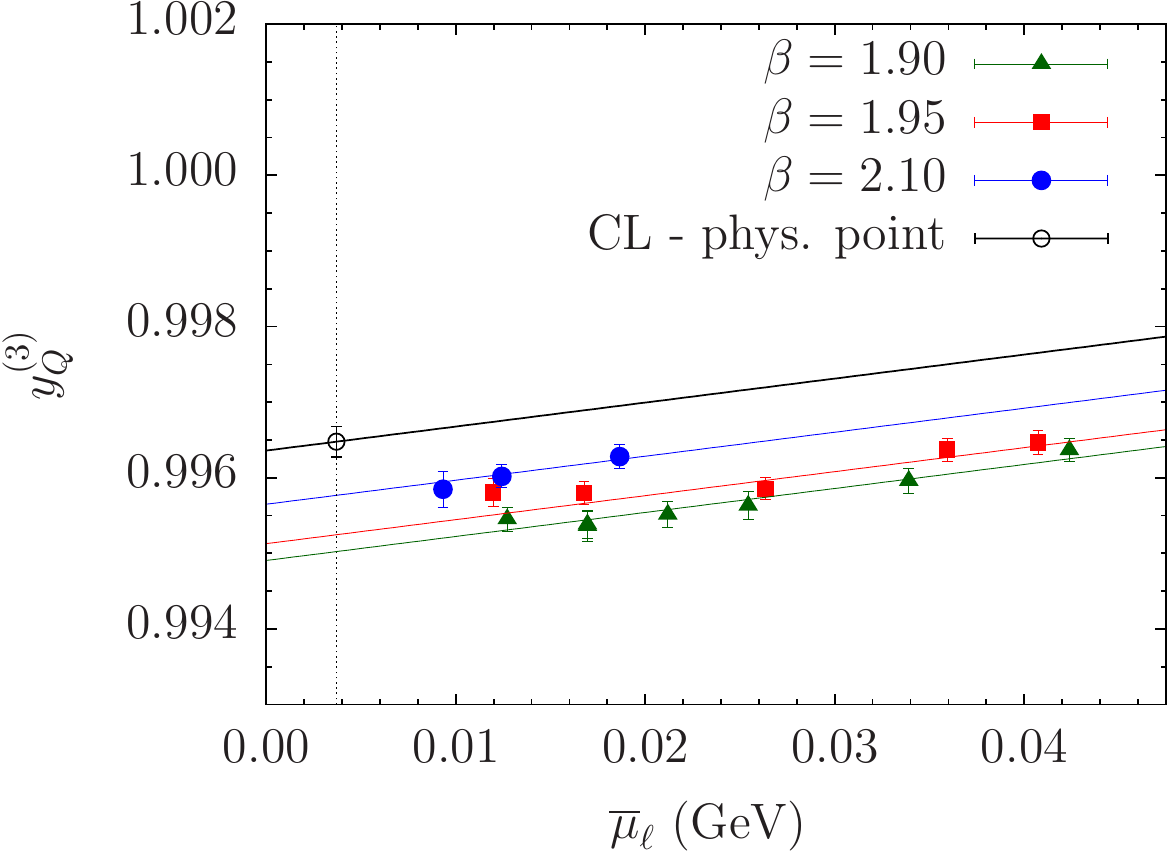} }}%
    \caption{Combined chiral and continuum fit of the ratio defined in Eq.~(\ref{eq:yn}) against the renormalised light
quark mass $\overline{\mu}_{\ell} = \overline{\mu}_{sea}$:   
(a) for the two largest values of heavy quark mass and (b) for intermediate values of heavy quark masses. 
The fit ansatz is linear both in $\overline{\mu}_{\ell}$
and in $a^2$.  
The empty black circle is our result at the physical $u/d$ quark mass point in the continuum limit. 
 }
    \label{fig:y8_y3}
\end{figure}

\begin{figure}%
    \centering
    \subfloat[]{{\includegraphics[width=7.5cm]{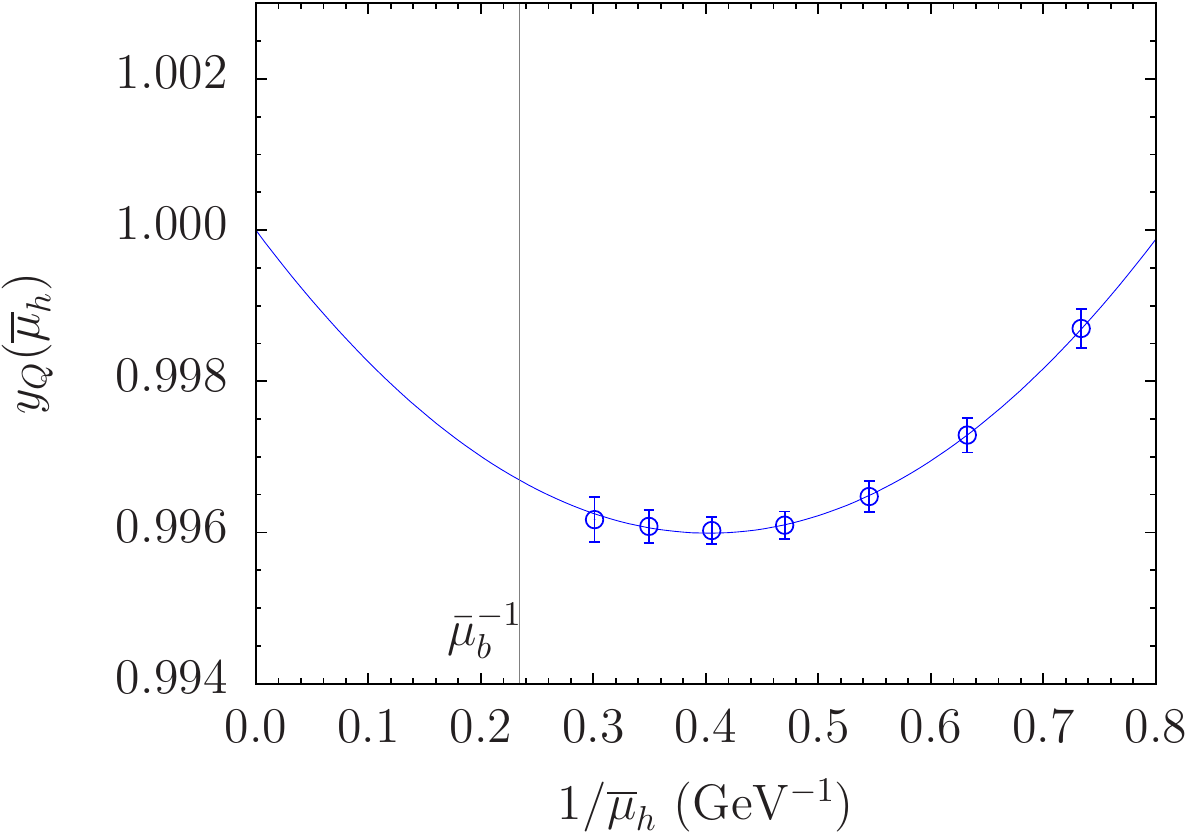} }}%
    \qquad
    \subfloat[]{{\includegraphics[width=7.5cm]{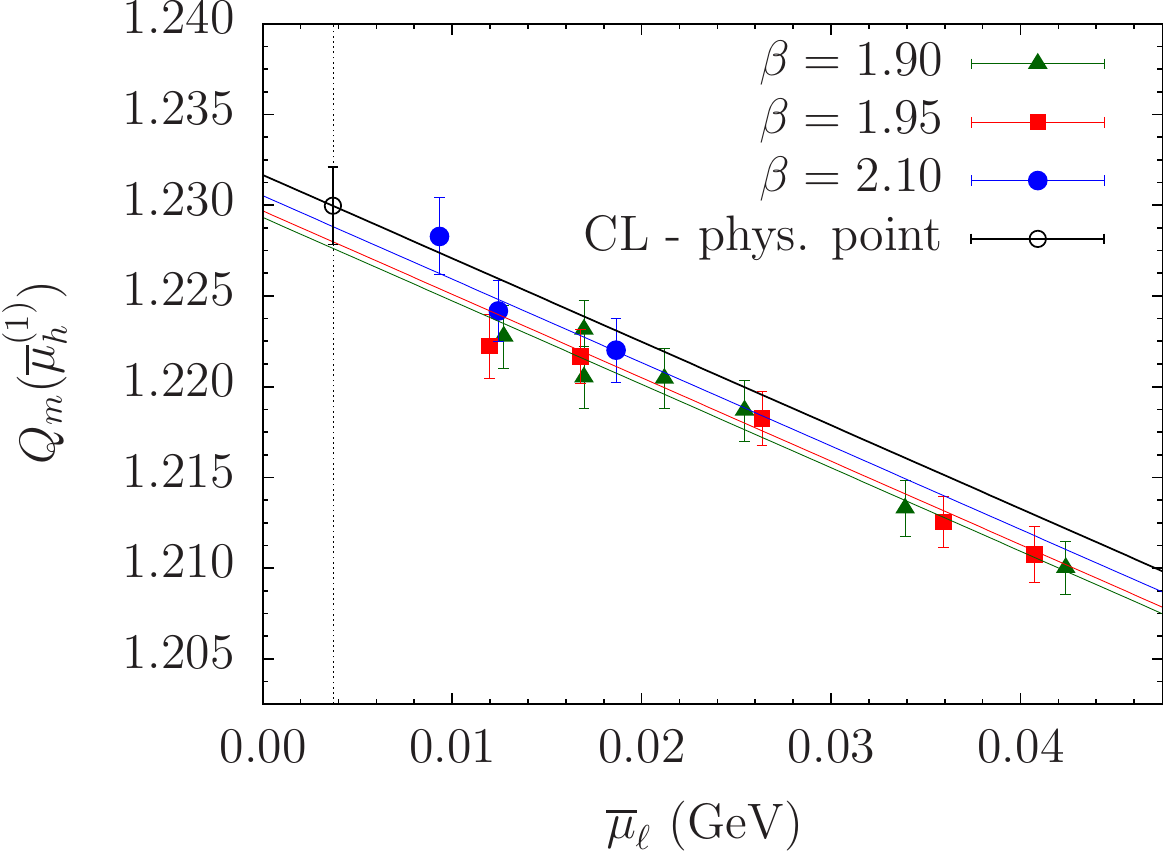} }}%
    \caption{(a) $y_{Q}(\overline{\mu}_h)$ against $1/\overline{\mu}_h$ using the fit ansatz of Eq.~(\ref{eq:y_vs_muh}). 
The vertical black thin line marks the position of $1/\overline{\mu}_b$. 
(b) Combined chiral and continuum fit at the triggering point {\it i.e.} for the quantity 
$Q_m(\overline\mu_h^{(1)})$ against the renormalised light quark mass $\overline\mu_{\ell}$. 
The empty black circle is our result at the physical $u/d$ quark mass point in the continuum
limit.  }
    \label{fig:y_vs_muh-trig-mass}
\end{figure}

At each value of the lattice spacing and sea quark mass ensemble we build the quantity 
\begin{equation}\label{eq:Qm}
Q_m \equiv \dfrac{M_{hs}}{(M_{h\ell})^{\gamma} (M_{cs})^{(1-\gamma)}},
 \end{equation} 
where $M_{hs}$ and $M_{h\ell}$ are the heavy-strange and 
heavy-light pseudoscalar masses, respectively, while we denote by $M_{cs}$ the mass of the pseudoscalar meson made out 
of a charm and a strange 
quark. The parameter $\gamma$, not subject to tuning, may take values, typically,  
in the interval $[0,1)$. 
We note that employing the dimensionless quantity $Q_{m}(\overline{\mu}_h)$ of  Eq.~(\ref{eq:Qm}) in our analysis 
we gain large cancellations of the lattice scale systematics on $m_b$.
Using HQET arguments we know that the asymptotic behaviour will be given by
\begin{equation}
\displaystyle \lim_{\mu_h^{\rm pole}\to \infty}  \left(\frac{M_{hs}/(M_{h\ell})^{\gamma}}{(\mu_h^{\rm pole})^{(1-\gamma)}
}\right)
= {\rm const.} ~, 
\end{equation}    
where $\mu_h^{\rm pole}$ is the heavy quark pole mass. We then consider a sequence of heavy quark 
masses\footnote{In the present analysis quark masses are 
expressed in the $\overline{{\rm MS}}$-scheme at the scale of $\mu = 2$ GeV.} such that any two successive masses have a 
common and fixed ratio i.e. 
$\overline{\mu}_h^{(n)} = \lambda \overline{\mu}_h^{(n-1)}$, $n = 2, 3,  \dots $. 
The next step is to construct at each value of the sea 
quark mass and  lattice spacing the following ratios:
\begin{eqnarray}
 y_Q(\overline\mu_h^{(n)},\lambda;\overline\mu_{\ell}, \overline\mu_s, a) &\equiv & 
\frac{Q_m(\overline\mu_h^{(n)};\overline\mu_\ell, \overline\mu_s, a)}{Q_m(\overline\mu_h^{(n-1)};
\overline\mu_\ell, \overline\mu_s, a)} 
\cdot
\left(\frac{\overline\mu^{(n)}_h \rho (\overline\mu^{(n)}_h,\mu)}
{\overline\mu^{(n-1)}_h \rho (\overline\mu^{(n-1)}_h,\mu)}\right)^{(\gamma - 1) } \nonumber \\
&=& \lambda^{(\gamma-1)} \frac{Q_m(\overline\mu_h^{(n)};\overline\mu_\ell, \overline\mu_s, a)}
{Q_m(\overline\mu^{(n)}_h/\lambda;\overline\mu_\ell, \overline\mu_s, a)}
\left(\frac{ \rho (\overline\mu^{(n)}_h,\mu)}{\rho( \overline\mu^{(n)}_h/\lambda,\mu)}\right)^{(\gamma - 1)}\,  
\label{eq:yn}
\end{eqnarray}   
with $n = 2, 3,  \dots $ and we have used the relation $\mu_{h}^{\rm{pole}} = \rho( \overline\mu_h,\mu)~\overline{\mu}_h(\mu)$  
between the $\overline{{\rm MS}}$ renormalised quark mass (at the scale $\mu$) and the pole quark mass.
The factors $\rho$'s are known perturbatively up to N$^3$LO~\cite{Chetyrkin:1999pq,Gray:1990yh, 
Broadhurst:1991fy, Chetyrkin:1999ys, Melnikov:2000qh }. For each pair of heavy quark masses we then 
carry out a simultaneous chiral and 
continuum fit of the quantity defined in Eq.~(\ref{eq:yn}) to obtain
$y_Q(\overline\mu_h) \equiv y_Q(\overline\mu_h,\lambda;\overline\mu_{u/d}, \overline\mu_s, a=0)$. 
By construction this quantity involves (double) ratios of 
pseudoscalar meson masses at successive values of the heavy quark mass, so  we expect that systematic uncertainties due to 
the use of the perturbative factors $\rho( \overline\mu_h,\mu)$ as well as discretisation errors will be 
quite suppressed\footnote{Notice that $M_{cs}^{(1-\gamma)}$ cancels out in the ratios defined in Eq.~(\ref{eq:yn}). The  
dependence on the scale $\mu$  in the determination of the factors  $\rho( \overline\mu_h,\mu)$ 
is also cancelled out in the ratios.}. 
In fact this is the case even for the largest values of the heavy quark mass used in this work as it can be seen in the plot 
of Fig.~\ref{fig:y8_y3}(a).  In Fig.~\ref{fig:y8_y3}(b)  the scaling behaviour of the ratios 
is shown at some intermediate value of heavy quark mass pair.  
Since in the quark mass ratios of Eq.~(\ref{eq:yn})    
we have taken account of the matching of QCD onto HQET, our ratio $y_Q(\overline\mu_h)$ 
has been defined such that the following ansatz is sufficient to describe the 
$\overline\mu_h$-dependence of $y_Q$~\footnote{For more details on this point see Appendix of Ref.~\cite{Dimopoulos:2011gx}.} 
\begin{equation}\label{eq:y_vs_muh}
y_Q(\overline\mu_h) = 1 + \frac{\eta_1}{\overline \mu_h} +  \frac{\eta_2}{{\overline \mu}_h^2}. 
\end{equation}  
In Eq.~(\ref{eq:y_vs_muh})  the constraint $\lim_{\overline\mu_h \to \infty} y_Q(\overline\mu_h) = 1$ has already been incorporated. 
This fit is illustrated in Fig.~\ref{fig:y_vs_muh-trig-mass}(a). 
Finally, we compute the $b$-quark mass through the {\it chain} equation 
\begin{equation}
y_Q(\overline\mu_h^{(2)})\, y_Q(\overline\mu_h^{(3)})\,\ldots \, y_Q(\overline\mu_h^{(K+1)})
 =\displaystyle \lambda^{K(\gamma - 1)} \,
\frac{Q_{m}(\overline\mu_h^{(K+1)})}{Q_{m}(\overline\mu_h^{(1)})} \cdot
\Big{(}\frac{\rho( \overline\mu_h^{(K+1)},\mu)}{\rho( \overline\mu_h^{(1)},\mu)}\Big{)}^{\gamma - 1} \label{eq:chain}
\end{equation}
in which the values of the factors in the (lhs) are evaluated using the result 
of the fit function ({\it viz.} Eq.~(\ref{eq:y_vs_muh})). 
The parameters $\lambda$, $K$ (integer) and $\overline{\mu}^{(1)}_h$
are such that  $Q_{m}(\overline\mu_h^{(K+1)})$  matches $(M_{Bs}/(M_B)^{\gamma}) (M_{Ds})^{(\gamma -1)}$, 
where $M_{Bs} = 5366.7(4)$ MeV, $M_B$ = 5279.3(3) MeV and $M_{Ds}$ = 1969.0(1.4) MeV 
are the experimental values of the $B_s$, $B$ and $D_s$ meson masses~\cite{Agashe:2014kda}, respectively. Moreover,   
$Q_{m}(\overline\mu_h^{(1)})$ (the so called {\it triggering point} of the chain equation) 
can be safely computed in the continuum limit and 
at the physical pion mass for any value  of $\overline\mu_h^{(1)}$ chosen in the region of the charm quark mass 
, see Fig.~\ref{fig:y_vs_muh-trig-mass}(b). The combined chiral and continuum fit ansatz we have used is linear 
in $\overline\mu_{\ell}$~\cite{Roessl:1999iu} and in $a^2$.

The result for the $b$-quark mass  
will be given by\footnote{Here the value for the $b$-quark mass is expressed in the $\overline{{\rm MS}}$-scheme 
at the scale of 2 GeV, {\it i.e.} the same scheme  and scale  we have decided to work in this analysis.} $\overline{\mu}_b = \lambda^{K}~ \overline{\mu}^{(1)}_h $. 
Figs.~\ref{fig:y8_y3} and \ref{fig:y_vs_muh-trig-mass} refer to one of the analyses we have performed in this work
where, by setting, $\overline{\mu}^{(1)}_h = 1.175$ GeV and $\gamma=0.75$, we find $(\lambda, ~ K) = (1.160, ~10)$.
We note here that for the running coupling  entering in the $\rho(\overline{\mu}_h, \mu)$ function 
we have used $\Lambda_{QCD}^{N_f=4}= 297(8)$ MeV~\cite{Agashe:2014kda}.  
 
\begin{table}[!h] 
\begin{center}
\scalebox{1.00}{
\begin{tabular}{|l|c|c|}
\hline
uncertainty (in \%)  & $m_{b}$ & $m_{b}/m_{c}$\tabularnewline
\hline
\hline
stat+fit & 0.9 & 0.7\tabularnewline
\hline
syst. discr. & 1.6 & 0.9\tabularnewline
\hline
syst. ratios   & 0.8 & 0.8\tabularnewline
\hline
syst. chiral  & 0.4 & 0.3\tabularnewline
\hline
syst. trig. point & - & 1.2 \tabularnewline
\hline
RI$'$ - $\overline{\text{MS}}$ matching  &  1.3 & -\tabularnewline
\hline
Total & 2.4 & 1.9  \tabularnewline
\hline
\end{tabular}
}
\caption{Full error budget for $m_b$ and $m_b/m_c$.  \label{Tab:budget_mb}} 
\end{center}
\end{table}
A detailed error budget is given in Table~\ref{Tab:budget_mb}. The description of the various entries follows:
\renewcommand\labelitemi{$\textendash$}
\begin{itemize}[noitemsep, leftmargin=*]
\itemsep0.2em
\item ``stat+fit": we gather the error coming from the statistical
uncertainties of correlators, the interpolation/extrapolation of the simulated quark masses
to the physical values, the extrapolation to the continuum limit, as well as the statistical
uncertainties of the RCs. We here recall that statistical errors have been evaluated using the jackknife method and 
fit cross correlations are taken into account by generating bootstrap samples for each gauge configuration ensemble.

\item ``syst. discr.": it refers to two sources of systematic uncertainty, both due to cutoff effects. 
The first is related 
to the two evaluations of the quark mass RC, called M1- and M2-type, 
which correspond to different ways in which the cutoff effects are treated in the RI-MOM calculation 
(see Appendix A of Ref.~\cite{Carrasco:2014cwa}). This amounts to about 1.4\%. 
The second one is the  difference (of about 0.8\%) between the result obtained through 
an analysis where data from the coarsest lattice spacing ($\beta=1.90$) have been excluded and the one that uses data from all three 
values of $\beta$. The two above systematic uncertainties have been added in quadrature. 

\item ``syst. ratios":   we collect five different types of systematic uncertainties added in quadrature: 
(a) systematic uncertainty,  of about 0.7$\%$, due to the choice of the parameter $\gamma$. In our analysis we have employed the 
following values for the parameter $\gamma = 0.0, 0.25, 0.50, 0.60, 0.75, 0.90$; 
(b) uncertainty in tuning the value of the step $\lambda$ to satisfy the chain equation (of about 0.3\%); 
(c) in our analysis we have made use of the NLL order formulae for the 
$\rho$'s while the use of LL or TL ones, thanks to the fact that we work with ratios, would lead to a discrepancy 
of about 0.3\% in the final results; (d) uncertainty of less than 0.1\% on the final result if we add to the fit ansatz of Eq.~(\ref{eq:y_vs_muh}) 
an extra cubic term in $1/\mu_h$;  (e) difference of the final result with the one obtained by excluding from 
the analysis the ratio corresponding to the heaviest quark mass pair (less than 0.1\%). 
Let us stress that the freedom of varying the value of the parameter $\gamma \in [0, 0.9]$ in our analysis\footnote{Notice that in practice 
for $ \gamma \in (0.9, 1.0)$ it becomes difficult to estimate the systematic uncertainty in the tuning 
of $\lambda$ from the chain equation~(\ref{eq:chain}).},   
at the cost of a moderate increase of the systematic error in the final value,   
allows to gain confidence  in estimating the systematic uncertainties due to discretisation 
effects and the use of the  fit ansatz given in Eq.~(\ref{eq:y_vs_muh}).       

\item ``syst. chiral": it refers to the systematic  uncertainty stemming from chiral extrapolation, which is  
estimated as the spread between the result obtained from all data and the one computed using data with pion mass smaller than 350 MeV.

\item ``RI$'$ - $\overline{\text{MS}}$ matching": for this systematic error estimate, concerning the matching between the two 
schemes at the typical scales the RCs are computed, we refer the reader to Appendix A of Ref.~\cite{Carrasco:2014cwa}.
\end{itemize}
The (small) experimental error (of about 0.01\% or less) on the values 
of the $B_{(s)}$ and $D_s$ pseudoscalar meson masses has a negligible impact on our error budget. 

Our final result for the $b$-quark mass 
is given by the average over the estimates obtained by varying the parameter $\gamma \in [0, 0.9]$ and using 
M1- or M2-type quark mass RC. 
The maximum half-difference between extreme values  
related to the investigation for each one of the sources of systematic error  is taken as our estimate of the 
corresponding systematic uncertainty. Systematic uncertainties are always added  in quadrature.
Finally, we get:
\begin{equation}\label{eq:mb}
m_b(\overline{{\rm MS}}, m_b)  = 4.26 (3)_{stat+fit} (10)_{syst} [10]~ {\rm  GeV}, 
\end{equation}    
where the total error (in brackets) is the sum in quadrature of the statistical and the systematic ones.

\subsubsection{Computation of $m_b/m_c$ and $m_b/m_s$}

The ratio method offers the advantage of determining the ratio $m_b/m_c$ in a simple and 
fully non-perturbative way. To this end we have to set the triggering point quark mass equal to the physical value 
of the charm quark mass, $\overline{\mu}_h^{(1)} = \overline{\mu}_c$. We then apply the ratio method employing the following quantity:
\begin{equation}\label{eq:Qrm}
\widehat{Q}_{m} = \dfrac{M_{hs}}{(M_{h\ell})^{\gamma} }.
 \end{equation}   
which, unlike the one defined in Eq.~(\ref{eq:Qm}), must be chosen dimensionful for ensuring charm scale dependence 
at the triggering point. So by implementing a similar procedure to the case of the $b$-quark mass it becomes possible to compute 
the $b$ to $c$ quark mass ratio directly from the relationship $\overline{\mu}_b = \lambda^{K} \overline{\mu}_c$.  
The error budget is also given in Table~\ref{Tab:budget_mb}. The various entries have a description similar to those for $m_b$. 
However there is an 
extra contribution under the name "syst. trig. point", which refer to  the systematic uncertainty related to residual  
uncertainties in the computation of the (dimensionful) quantity $\widehat{Q}_{m}$ at the triggering point. 
These uncertainties are not related (directly) to the scale setting and to renormalisation contant's 
uncertainties that in the  bottom to charm quark mass ratio  clearly cancel out. They include instead the following systematic 
uncertainties related to the determination of $m_c$ (see Ref.~\cite{Carrasco:2014cwa}) which refer 
to the two choices of scaling variable in that fit analysis, 
the systematic chiral and discretisation uncertainties, the systematic uncertainty stemming from the matching to $M_D$ and $M_{Ds}$ 
as well as statistical uncertainties in the pseudoscalar mass values computed in the charm region. 
We consider the sum in quadrature of the above uncertainties 
to get our estimate.

Our final result reads:
\begin{equation}\label{eq:mbmc}
m_b / m_c  = 4.42 (3)_{stat+fit} (8)_{syst} [8],
\end{equation} 
where the total error (in brackets) is the sum in quadrature of the statistical and  systematic ones. 
As stated above in the quark mass ratio computation the uncertainties due to  the RC and  
renormalisation scheme  as well as the systematic lattice scale uncertainties cancel out.

Finally, by combining the result of Eq.~(\ref{eq:mbmc}) with the result  
\begin{equation}\label{eq:mcms}
m_c/m_s = 11.62(16)_{stat+fit}(1)_{syst} [16]
\end{equation} 
presented in Ref.~\cite{Carrasco:2014cwa} we obtain the value for the bottom to strange quark mass ratio: 
\begin{equation}\label{eq:mbms}
m_b/m_s = 51.4 (1.1)_{stat+fit} (0.9)_{syst} [1.4],
\end{equation}
where, again, the sum of the statistical and the systematic errors in quadrature give the the total error (in brackets).  
The error estimate has been obtained assuming full correlation between the ``stat+fit" uncertainties 
of Eqs~(\ref{eq:mbmc}) and (\ref{eq:mcms}), which thus have been added linearly, whereas
systematic uncertainties have been added in quadrature. 
Our result of Eq.~(\ref{eq:mbms}) compares well with the (non-perturbative) result $m_b/m_s = 52.55(55)$ 
obtained by the HPQCD collaboration~\cite{Chakraborty:2014aca}. 
It is also in agreement with the Georgi-Jarlskog prediction~\cite{Georgi:1979df} 
that for certain classes of grand unified theories the ratio of $b$ to $s$
quark masses should be equal to $3\, m_{\tau} / m_{\mu} = 50.45$.


\subsection{B-pseudoscalar decay constants}

\begin{figure}[!h]
    \centering
    \subfloat[]{{\includegraphics[width=7.5cm]{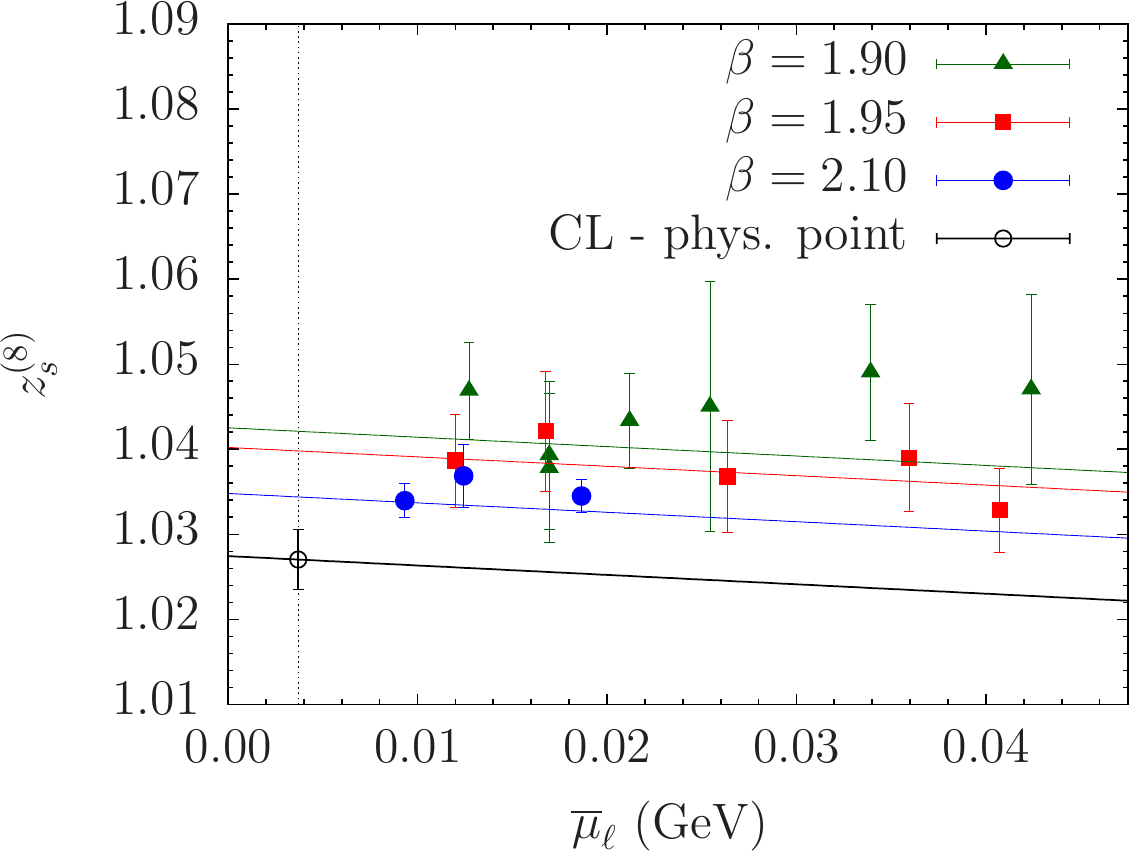} }}%
    \qquad
    \subfloat[]{{\includegraphics[width=7.5cm]{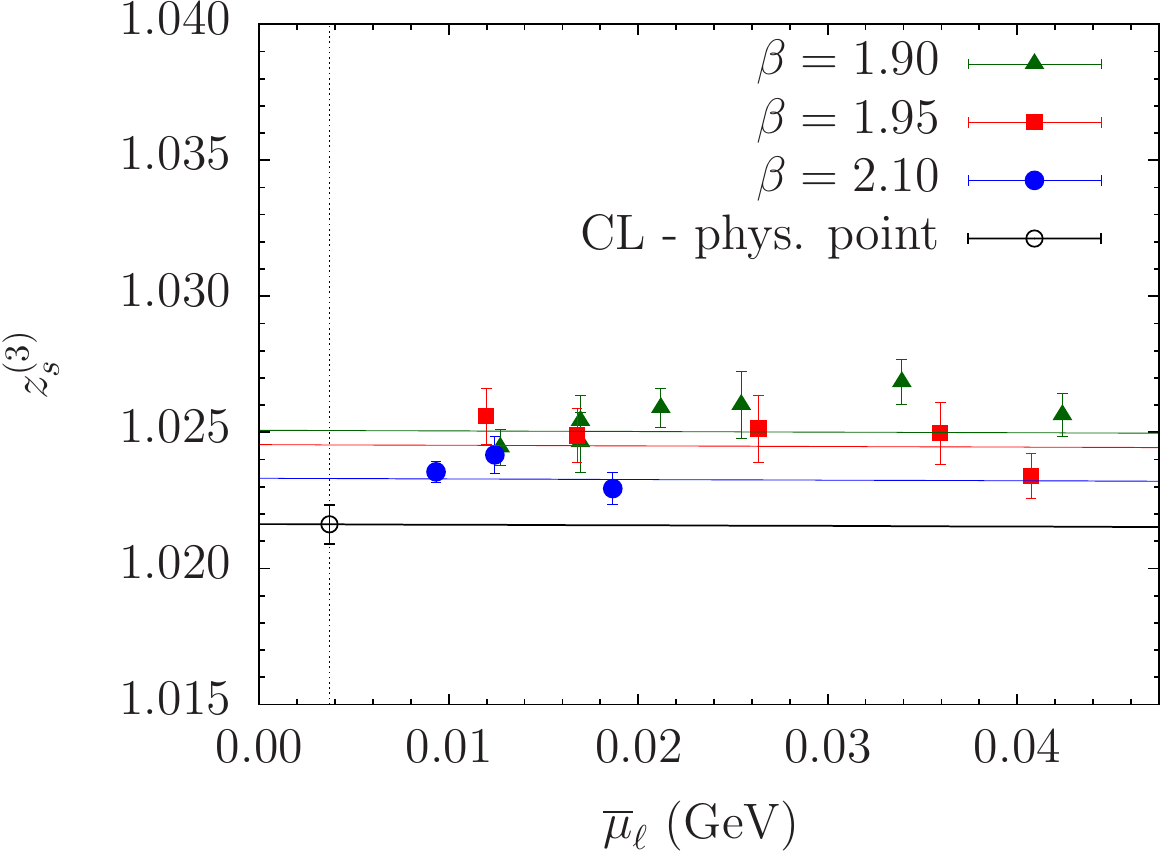} }}%
    \caption{Combined chiral and continuum fit for the ratio
$z_{s}$ against  $\overline\mu_{\ell}$ calculated: (a) between the two largest heavy quark mass values used in this work; 
(b) for  intermediate values of the heavy quark masses. 
The empty black circle is our result at the physical $u/d$ quark mass point in the continuum limit.}
    \label{fig:zs_eq_8_eq_3}
\end{figure}

\begin{figure}[!h]
    \centering
    \subfloat[]{{\includegraphics[width=7.5cm]{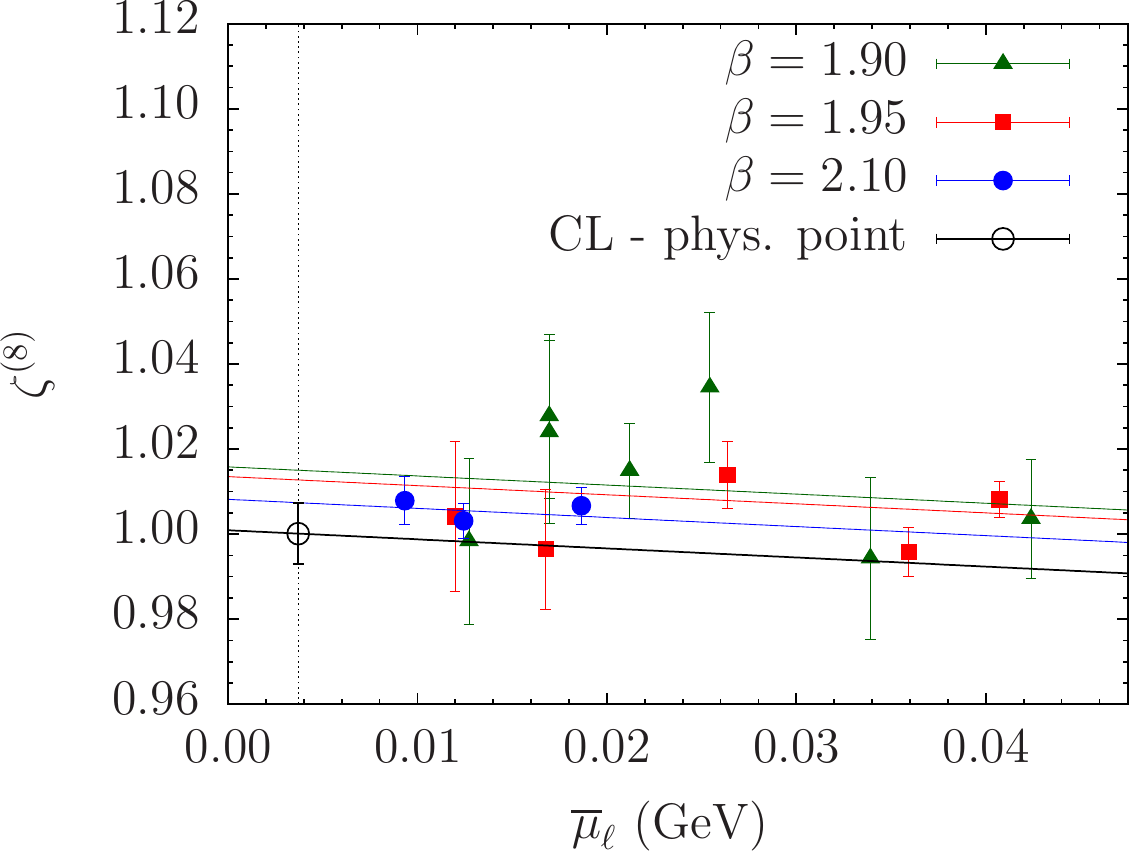} }}%
    \qquad
    \subfloat[]{{\includegraphics[width=7.5cm]{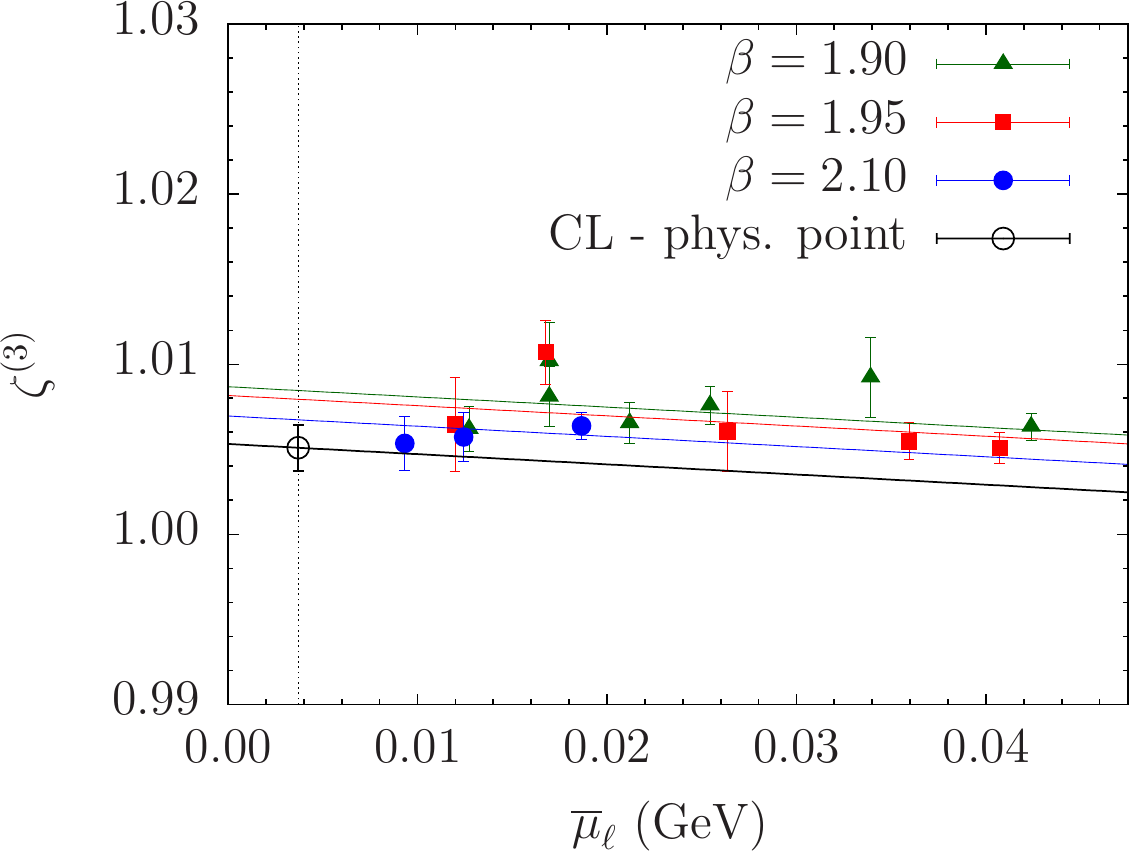} }}%
    \caption{Same as in Fig.~\ref{fig:zs_eq_8_eq_3} for the ratio $\zeta$. }
    \label{fig:zeta_eq_8_eq_3}
\end{figure}

At each value of the lattice spacing and sea quark mass ensemble we evaluate the quantity  
\begin{equation}\label{eq:fhs}
\mathcal{F}_{hq} \equiv f_{hq} / M_{hq}, ~~ q = \ell, s
\end{equation}
for which the appropriate HQET asymptotic conditions lead to 
\begin{equation}
\displaystyle \lim_{\mu_h^{\rm pole}\to \infty}  \mathcal{F}_{hq} ~(\mu_h^{\rm pole})^{3/2}
= {\rm const.} ~, 
\end{equation}  
and 
\begin{equation}
\lim_{\mu_h^{\rm{pole}}\to \infty} \Big(\mathcal{F}_{hs}/\mathcal{F}_{h\ell}\Big) = \mbox{const.}, 
\end{equation}
Based on QCD to HQET matching of heavy-light meson decay constant and quark mass we define the ratios
\begin{eqnarray}
z_s(\overline\mu_h,\lambda;\overline\mu_s, a) &=&
\lambda^{3/2} \frac{\mathcal{F}_{hs}(\overline\mu_h, \overline\mu_s, a)}
{\mathcal{F}_{hs}(\overline\mu_h/\lambda, \overline\mu_s, a)}
\cdot \frac{C^{stat}_A(\mu^*,\overline\mu_h/\lambda)}{C^{stat}_A(\mu^*,\overline\mu_h)}
\frac{[\rho( \overline\mu_h,\mu)]^{3/2}}{[\rho( \overline\mu_h/\lambda,\mu)]^{3/2}} \, \\
z_{d}(\overline\mu_h,\lambda;\overline\mu_\ell, a) &=&
\lambda^{3/2} \frac{\mathcal{F}_{h\ell}(\overline\mu_h,\overline\mu_\ell, a)}
{\mathcal{F}_{h\ell}(\overline\mu_h/\lambda,\overline\mu_\ell, a)}
\cdot \frac{C^{stat}_A(\mu^*,\overline\mu_h/\lambda)}{C^{stat}_A(\mu^*,\overline\mu_h)}
\frac{[\rho( \overline\mu_h,\mu)]^{3/2}}{[\rho( \overline\mu_h/\lambda,\mu)]^{3/2}} \, 
\label{eq:z_zs_ratios}
\end{eqnarray}
The factor $C^{stat}_A(\mu^*,\overline\mu_h)$ is  
known up to N$^2$LO in PT~\cite{ChetGrozin}. It provides  
the matching between the $(h\ell)$ decay constant in QCD
and its static-light counterpart in HQET\footnote{Notice that the  renormalization scale $\mu^*$ 
of HQET as well as the quark mass renormalization scale $\mu$ cancel when ratios are considered.}.
For the calculation of the decay constant ratio we also form the double ratio
\begin{equation} \label{eq:zeta}
\zeta(\overline\mu_h,\lambda;\overline\mu_\ell, \overline\mu_s, a) =\dfrac{z_s(\overline\mu_h,\lambda; \overline\mu_s, a)}
{z_d(\overline\mu_h,\lambda;\overline\mu_\ell, a)} .
\end{equation}
The ratios $z_d$, $z_s$ and $\zeta$    
have by construction an exactly known static limit equal to unity. They also show smooth  
chiral and continuum combined behavior. This is a consequence of the fact that $z_d$, $z_s$ and $\zeta$ (as it is also the case for 
the $y$ ratios) are simply ratios 
of quantities evaluated at nearby values of the heavy quark mass for which discretisation errors get suppressed. 
Figs \ref{fig:zs_eq_8_eq_3}(a) and \ref{fig:zeta_eq_8_eq_3}(a) are two examples illustrating the quality of the combined 
chiral and continuum fits for $z_s$ and $\zeta$ respectively, 
at the largest heavy quark mass values used in the decay constant analysis. 
See also the analogous Figs~\ref{fig:zs_eq_8_eq_3}(b) and \ref{fig:zeta_eq_8_eq_3}(b)
for the same quantities at intermediate values of the heavy quark mass pair.  Notice that,  
since the cutoff effects for the ratios are under good control, even for rather 
large values of the heavy quark mass pairs the combined chiral and continuum fits of ratios are reliable.

\begin{figure}[!h]
    \centering
    \subfloat[]{{\includegraphics[width=7.5cm]{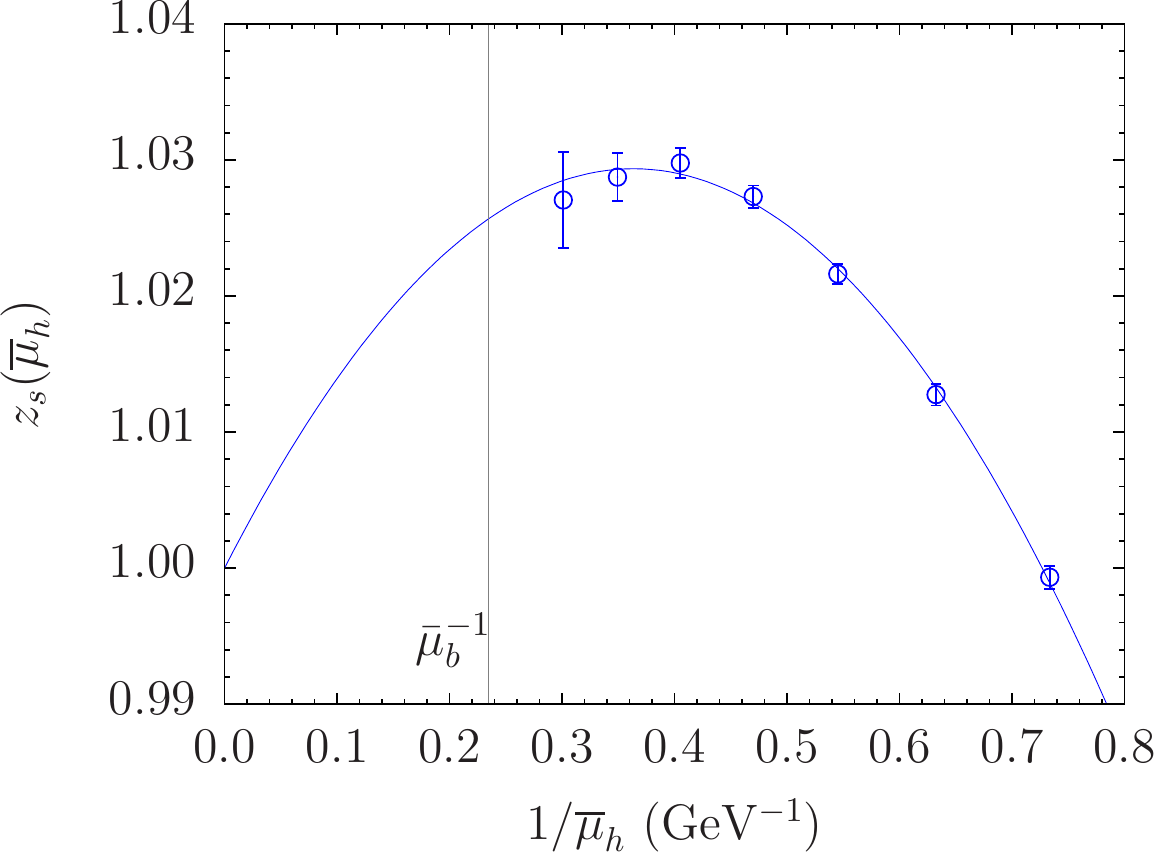} }}%
    \qquad
    \subfloat[]{{\includegraphics[width=7.5cm]{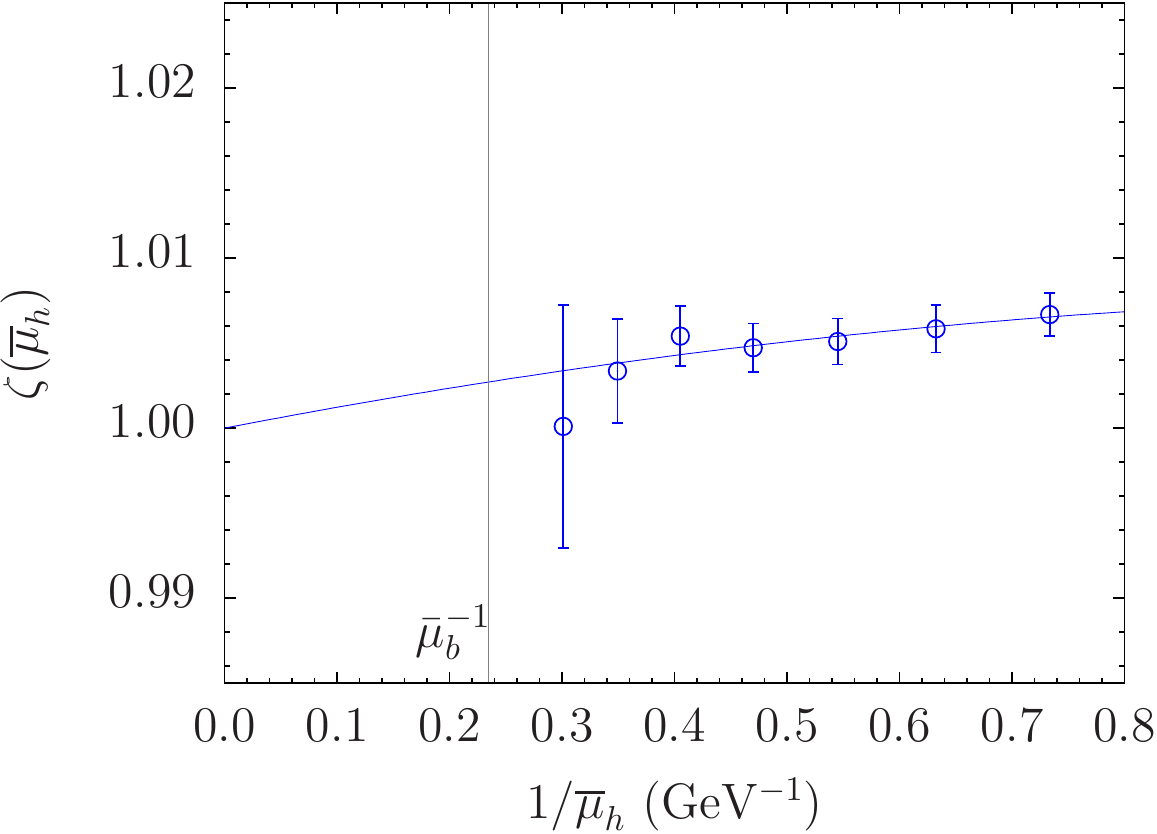} }}%
    \caption{Fit of $z_s(\overline\mu_h)$ (left panel) and  $\zeta(\overline\mu_h)$ (right panel) against $1/\overline\mu_h$. 
The fit function in both panels has a polynomial form of the type given in  Eq.~(\ref{eq:y_vs_muh}). 
 The vertical black thin line  marks the position of $1/\overline\mu_b$. }
    \label{fig:zs-zeta_vs_muh}
\end{figure}

In Figs~\ref{fig:zs-zeta_vs_muh}(a) and \ref{fig:zs-zeta_vs_muh}(b) we show the dependence of $z_s(\overline\mu_h)$ 
and $\zeta(\overline\mu_h)$ on the inverse heavy quark mass, respectively. The fit ans\"atze  we have 
used are polynomial fit functions in the inverse heavy quark mass analogous to the one displayed in Eq.~(\ref{eq:y_vs_muh}). 
For the case of the double ratio 
$\zeta(\overline\mu_h)$ we have also tried a linear fit in $1/\overline\mu_h$ 
where the exact condition $\lim_{\overline{\mu}_h \rightarrow \infty} \zeta(\overline\mu_h) = 1$ is explicitly implemented.  

In order to determine $f_{Bs}$ and $f_{Bs}/f_{B}$ we exploit the equations 
\begin{eqnarray} \label{eq:zs_chain}
 \hspace*{-0.5cm} z_s(\overline\mu_h^{(2)})\, z_s(\overline\mu_h^{(3)})\,\ldots \, z_s(\overline\mu_h^{(K+1)}) &=& \lambda^{3 K/2} \,
\frac{\mathcal{F}_{hs}(\overline\mu_h^{(K+1)})}{\mathcal{F}_{hs}(\overline\mu_h^{(1)})} \cdot
 \frac{C^{stat}_A(\mu^*, \overline\mu_h^{(1)})}{C^{stat}_A(\mu^*, \overline\mu_h^{(K+1)})}
\left(\frac{\rho( \overline\mu_h^{(K+1)},\mu)}{\rho( \overline\mu_h^{(1)},\mu)}\right)^{3/2} ,  \\
 \hspace*{-0.5cm} \zeta(\overline\mu_h^{(2)})\, \zeta(\overline\mu_h^{(3)})\,\ldots \, \zeta(\overline\mu_h^{(K+1)}) &=& 
\left( \frac{\mathcal{F}_{hs}(\overline\mu_h^{(K+1)})/\mathcal{F}_{hu/d}(\overline\mu_h^{(K+1)})}{\mathcal{F}_{hs}(\overline\mu_h^{(1)})
/\mathcal{F}_{hu/d}(\overline\mu_h^{(1)})} 
\right) . 
\end{eqnarray}

\begin{figure}[!h]
    \centering
    \subfloat[]{{\includegraphics[width=7.5cm]{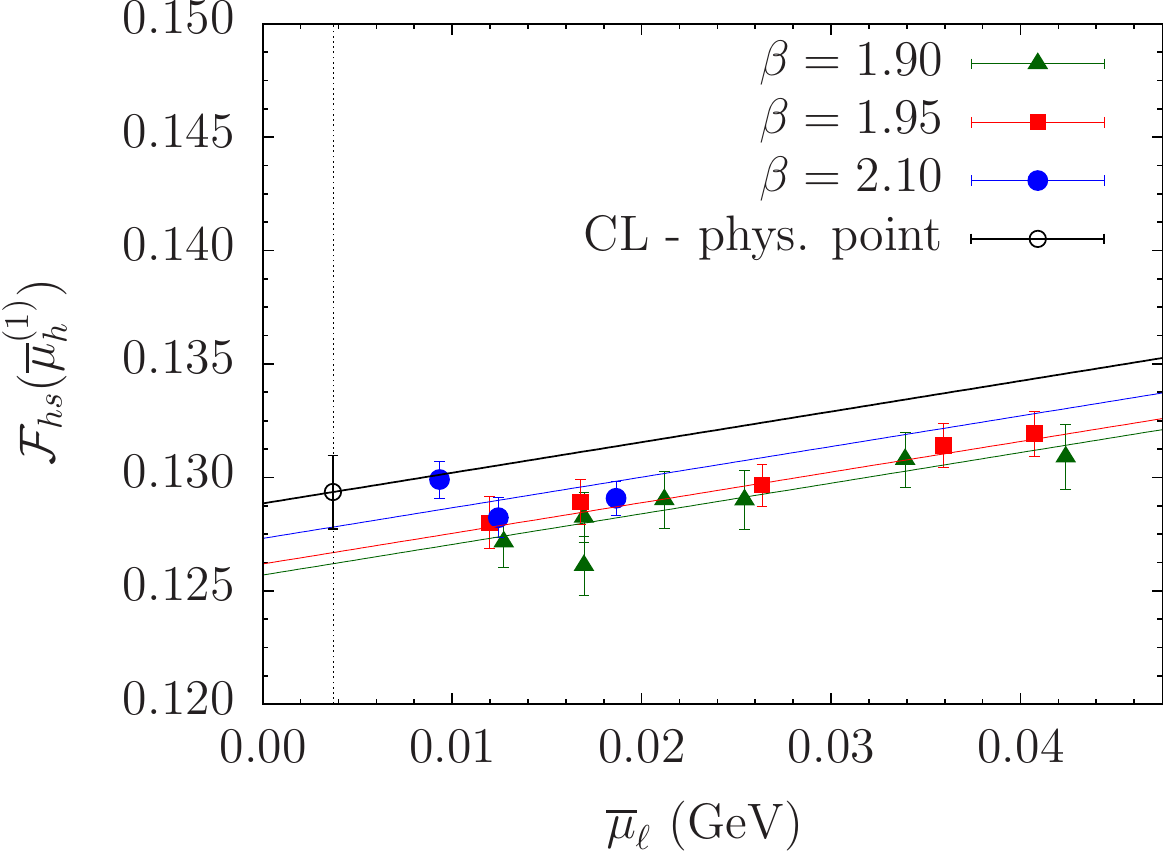} }}%
    \qquad
    \subfloat[]{{\includegraphics[width=7.5cm]{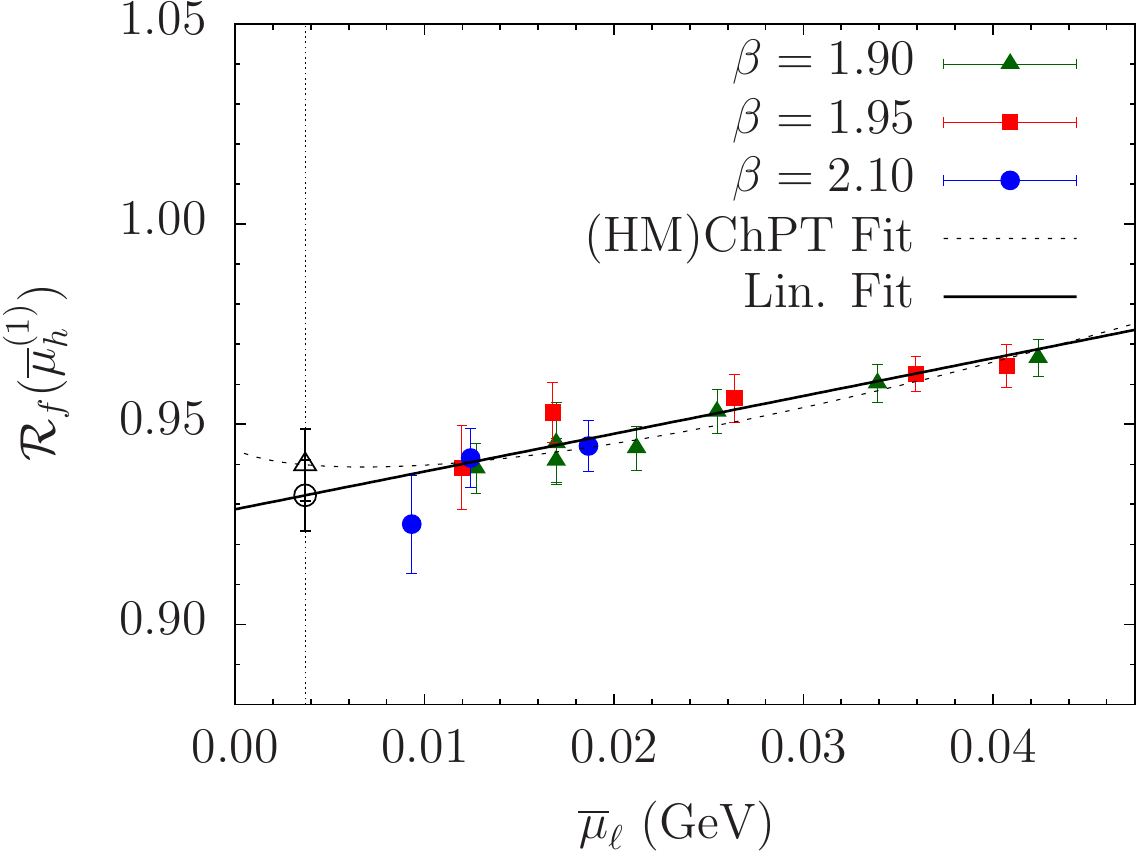} }}%
    \caption{Combined chiral and continuum fit against $\overline\mu_{\ell}$: 
    (a) linear fit in $\overline{\mu}_{\ell}$ and in $a^2$ for the triggering point of 
$\mathcal{F}(\overline{\mu}_h^{(1)})$; 
(b) fit ans\"atze for  ${\cal R}_f(\overline{\mu}_h^{(1)})$ given in Eqs~(\ref{eq:lin_ans}) and (\ref{eq:HMChPT_ratiof}). 
The empty black symbols denote results at the physical $u/d$ quark mass point in the continuum
limit.  }
    \label{fig:fhs-zeta_trig}
\end{figure}

The values of the l.h.s.'s of the above equations are taken from  the  fits of Figs~\ref{fig:zs-zeta_vs_muh}(a) 
and \ref{fig:zs-zeta_vs_muh}(b), respectively. 
Setting $\overline\mu_h^{(K+1)}=\overline\mu_b$ and having determined the values of 
$\mathcal{F}_{hs}(\overline\mu_h^{(1)})$ and $[\mathcal{F}_{hs}(\overline\mu_h^{(1)})/\mathcal{F}_{hu/d}(\overline\mu_h^{(1)})]$ from a 
combined chiral and continuum fit, and using experimental input for the  $B_s$ and $B$-meson masses, we finally 
obtain our results for $f_{Bs}$ and $(f_{Bs}/f_{B})$. The use of the observable of Eq.~(\ref{eq:fhs}) 
yields the continuum limit determination of $f_{Bs}$ in physical units via the experimental value of $M_{Bs}$, leading thus to the  
elimination of the lattice scale systematic uncertainty. 
As for the combined chiral and continuum fit of the triggering point quantity $\mathcal{F}_{hs}(\overline\mu_h^{(1)})$, this  
poses no problems because ${\cal F}_{hs}(\overline{\mu}_h^{(1)})$ exhibits only tolerably small cutoff 
effects and very weak dependence on the light quark mass (see Fig.~\ref{fig:fhs-zeta_trig}(a)).

In order to estimate the triggering point ratio
$[\mathcal{F}_{hs}(\overline\mu_h^{(1)})/\mathcal{F}_{hu/d}(\overline\mu_h^{(1)})]$  we build the following double ratio  
\begin{equation}
{\cal R}_f = [(\mathcal{F}_{hs}/\mathcal{F}_{h \ell}) / (f_{s \ell}/f_{\ell \ell})] 
\end{equation}
which provides the advantage of large cancellations in the 
chiral logarithmic terms~\cite{Becirevic:2002mh, Blossier:2009bx}.  
One then can get the desired triggering point ratio by combining  
the continuum limit result for ${\cal R}_f$  with the analogous result for the
ratio of the $K$ to $\pi$ decay constants, $(f_K/f_{\pi})$.  
In Fig.~\ref{fig:fhs-zeta_trig}(b) we present the combined continuum and chiral extrapolation for ${\cal R}_f$. 
We have used two fit ans\"atze. The first  is linear in 
$\overline\mu_{\ell}$ while the second one  is suggested by the combined use of the SU(2) ChPT and 
HMChPT. They read
\begin{eqnarray}\label{eq:zeta_trig}
 {\cal R}_f^{(1)} &=& a_h^{(1)} + b_h^{(1)} \overline\mu_{\ell} + D_h^{(1)} a^2 \label{eq:lin_ans}\\
 {\cal R}_f^{(2)}  &=& a_h^{(2)}\Big[1 + b_h^{(2)} \overline\mu_{\ell} + \Big[ \frac{3(1+3\hat{g}^2)}{4} - \frac{5}{4} \Big]
\frac{2 B_0 \overline\mu_{\ell}}{(4 \pi f_0)^2} 
{\rm log}\Big( \frac{2 B_0 \overline\mu_{\ell}}{(4 \pi f_0)^2}  \Big)\Big]  + D_h^{(2)} a^2.  \label{eq:HMChPT_ratiof} 
\end{eqnarray}
The magnitude  of the logarithmic term in this fit depends on the value of $\hat{g}$. 
Given the form of Eq.~(\ref{eq:HMChPT_ratiof}) we have used $\hat{g}=0.61(7)$~\cite{Agashe:2014kda} since for this value we get the most   
conservative estimate for the fit systematic uncertainty.

As it can be noticed from Fig.~\ref{fig:fhs-zeta_trig}(b) discretisation effects on ${\cal R}_f$ are small. 
Moreover, the two estimates for the triggering point ratio 
at the physical light quark mass are compatible within less than two standard deviations. 
So we take their average as our best estimate and 
we consider their half difference as a systematic uncertainty.

The central values of $f_{Bs}$ and $f_{Bs}/f_B$ have been obtained from the weigthed average over the various estimates 
corresponding to the sets of values ($\overline{\mu}_h^{(1)}, \lambda, K, \gamma)$   
employed in our $b$-quark mass  analysis. 
The $f_B$ computation is carried out  through the expression $f_B = f_{Bs} / (f_{Bs}/f_{B})$.  We now give the description
of the full error budget for the decay constants, presented in Table~\ref{Tab:budget_f}:  
\renewcommand\labelitemi{$\textendash$}
\begin{itemize}[noitemsep, leftmargin=*]
\itemsep0.2em
\item ``stat+fit": this has been estimated along the same lines as for the $b$-quark mass. 

\item ``syst. discr.": it includes two sources of systematic discretisation errors, then added in quadrature.  
Concerning the first one we take into account the fact that for the decay constants computation at the $b$-quark point 
we have collected as many estimates as there are the respective sets   
($\overline{\mu}_h^{(1)}, \lambda, K, \gamma)$ employed in 
our $b$-mass analysis. We then consider the maximum spread of these results from their average (about 0.5\% 
for $f_{Bs}$ and 0.4\% for $f_{Bs}/f_B$).
The second systematic error related to cutoff effects has been estimated by investigating the impact of removing from 
our analysis the coarsest lattice ($\beta = 1.90$) data. 
The maximum difference between results from the full data analysis and the 
one when only  data from the two finest lattices are used amounts to 1.2\% for $f_{Bs}$ and 0.4\% for $f_{Bs}/f_B$.

\item ``syst. ratios": we have checked the impact on our final results  of the various sources of 
systematic uncertainty related to the ratio analysis. We have worked along the same lines as for the 
$b$-mass error budget. In particular we have checked the effects by (a) varying the polynomial fit ansatz 
used for interpolating to the $b$-quark mass, (b) excluding the heaviest quark mass pair from our analysis 
and (c) changing the perturbative 
order for the $\rho$'s  and $C_A^{stat}$ from NLL to LL. None of these tests gave a change to the values of 
$f_{Bs}$ and $f_{Bs}/f_B$ larger than 0.3-0.4\%. 
The final estimates in Table~\ref{Tab:budget_f} correspond to the sum in quadrature of the individual spreads due to (a), (b) and (c).

\item ``syst. chiral": we estimate the systematic uncertainty due to chiral extrapolation 
from the difference between results obtained from all data or using only data corresponding to pion mass less than 350 MeV. 

\item ``syst. trig. point $\&$ $f_{K}/f_{\pi}$": this concerns only $f_{Bs}/f_B$ and $f_B$ and it is given as  
the sum in quadrature of the chiral extrapolation systematic 
uncertainty, which we estimate from the spread of results obtained from the two fit ans\"atze of 
Eqs~(\ref{eq:lin_ans}) and (\ref{eq:HMChPT_ratiof}) (of about $0.4\%$),  
and the error in the determination of $f_K/f_{\pi}$; for the latter we have used the value 
$f_{K}/f_{\pi} = 1.188(15)$ from Ref.~\cite{Carrasco:2014poa}.  
\end{itemize}

\begin{table}[!h] 
\begin{center}
\scalebox{1.00}{
\begin{tabular}{|l|c|c|c|}
\hline
uncertainty (in \%)  & $f_{Bs}$ & $f_{Bs}/f_{B}$ & $f_B$ \tabularnewline
\hline
\hline
stat+fit & 1.7 & 1.5  & 2.5\tabularnewline
\hline
syst.  discr.  & 1.3 & 0.6 & 0.7\tabularnewline
\hline
syst. ratios   & 0.5 & 0.3 & 0.6\tabularnewline
\hline
syst. chiral & 0.3 & 0.2 & 0.4\tabularnewline
\hline
syst. trig. point $\&$ $f_{K}/f_{\pi}$ & - & 1.3 & 1.3\tabularnewline
\hline
Total & 2.2 & 2.1  & 3.0\tabularnewline
\hline
\end{tabular}
}
\caption{Full error budget for $f_{Bs}$, $f_{Bs}/f_{B}$ and $f_B$. \label{Tab:budget_f} } 
\end{center}
\end{table}
\noindent Our  final results for the decay constants read:
\begin{eqnarray}
f_{Bs} &=& 229 (4)_{stat+fit} (3)_{syst}[5] ~{\rm MeV}, \label{eq:fB1} \\ 
f_{Bs}/f_B &=& 1.184 (18)_{stat+fit} (18)_{syst} [25], \label{eq:fB2} \\ 
f_B &=& 193 (5)_{stat+fit} (3)_{syst} [6] ~{\rm MeV}, \label{eq:fB3} \\ 
\left( f_{Bs}/f_B\right) / \left( f_K / f_{\pi}\right) &=& 0.997(15)_{stat} (7)_{syst} [17] \label{eq:fB}
\end{eqnarray}
where the total error (in brackets) is the sum in quadrature of the statistical and the systematic ones.

\section{Conclusions}
\label{sec:concl}
 
\begin{figure}[!h]%
    \centering
    \subfloat[]{{\includegraphics[width=6.0cm]{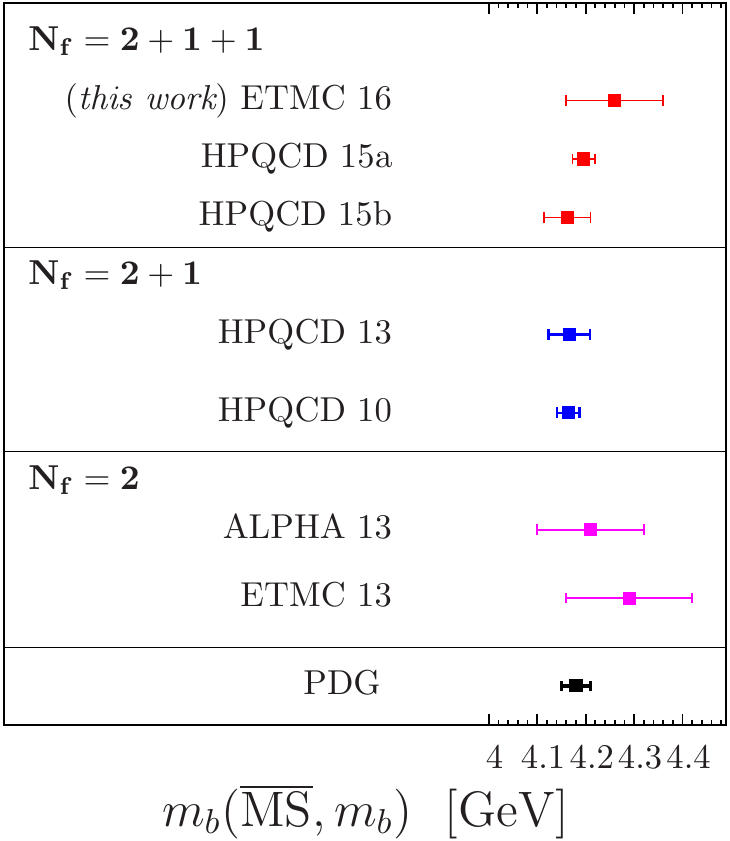} }}%
    \qquad
    \subfloat[]{{\includegraphics[width=6.0cm]{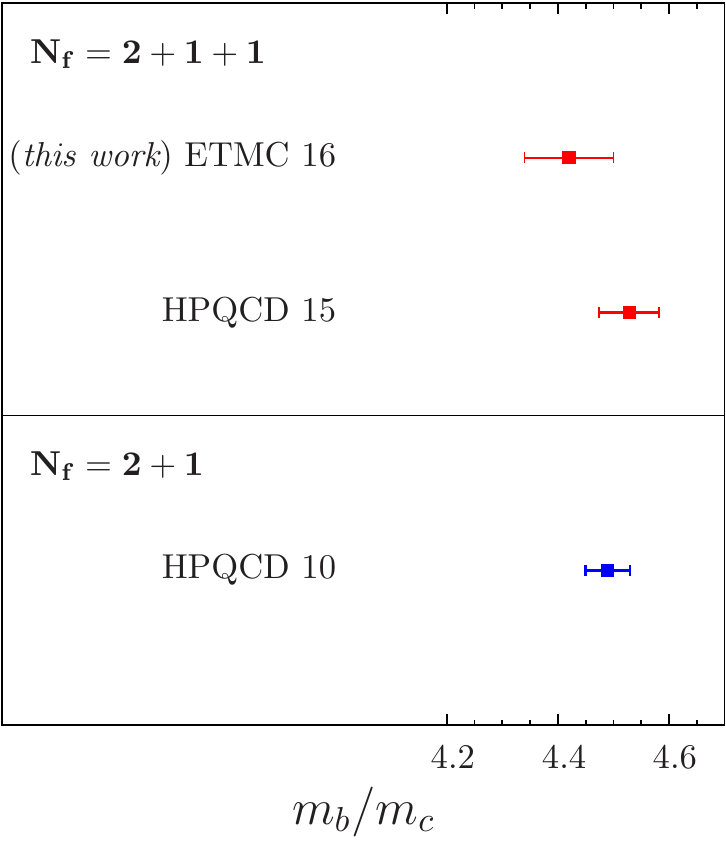} }}%
    \caption{A comparison of the available continuum extrapolated lattice determinations for $m_b$, panel (a), and $m_b/m_c$, panel (b). 
    For the other results we refer to (from top to bottom) 
    (a) Refs.~\cite{Colquhoun:2014ica, Chakraborty:2014aca},~\cite{Lee:2013mla},~\cite{McNeile:2010ji},~\cite{Bernardoni:2013xba},
    ~\cite{Carrasco:2013zta},~\cite{Agashe:2014kda}; 
    (b) Refs.~\cite{Chakraborty:2014aca},~\cite{McNeile:2010ji}.    
}
    \label{fig:compar_mass}
\end{figure}

\begin{figure}[!h]%
    \centering
    \subfloat[]{{\includegraphics[width=6.0cm]{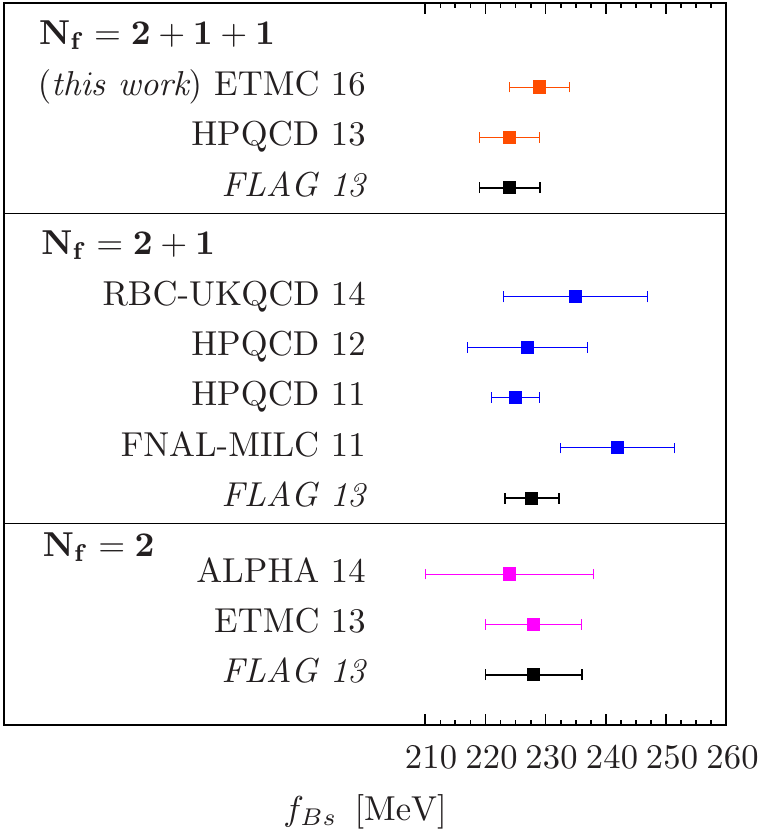} }}%
    \qquad
    \subfloat[]{{\includegraphics[width=6.0cm]{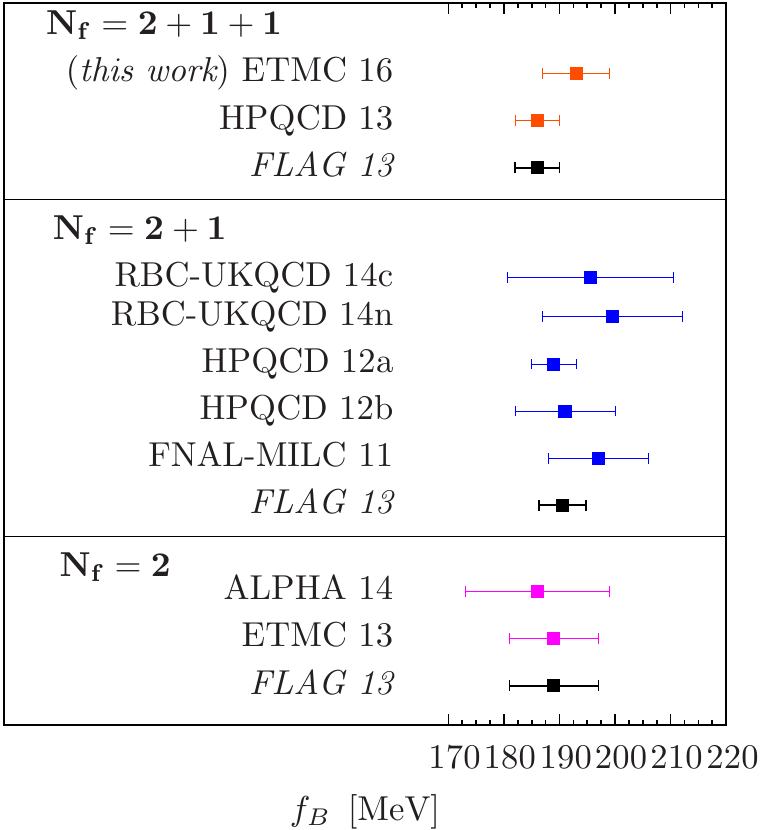} }}%
    \qquad,
    \subfloat[]{{\includegraphics[width=6.0cm]{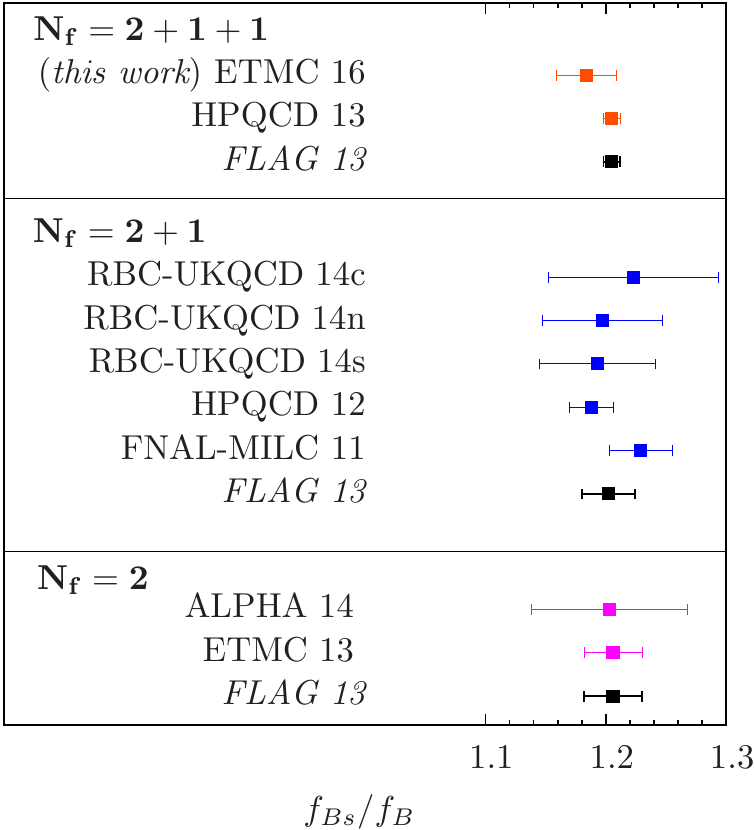} }}%
    \caption{A comparison of the available continuum extrapolated lattice determinations for $f_{Bs}$, panel (a),  $f_B$, panel (b) 
    and $f_{Bs}/f_B$, panel (c).  For results by other groups we refer to (from top to bottom) (a) Refs.~\cite{Dowdall:2013tga},
    ~\cite{Aoki:2013ldr},~\cite{Christ:2014uea},~\cite{Na:2012kp},~\cite{McNeile:2011ng},~\cite{Bazavov:2011aa},~\cite{Aoki:2013ldr},~\cite{Bernardoni:2014fva},~\cite{Carrasco:2013zta},~\cite{Aoki:2013ldr};
    {\it FLAG 13} estimates are determined by HPQCD 13 for $N_f=2+1+1$, HPQCD 12, HPQCD 11 and FNAL-MILC 11 for $N_f=2+1$ and ETMC 13 for 
    $N_f=2$;    
    (b) Refs.~\cite{Dowdall:2013tga},~\cite{Aoki:2013ldr},~\cite{Christ:2014uea},~\cite{Christ:2014uea},~\cite{Na:2012kp},~\cite{Na:2012kp},~\cite{Bazavov:2011aa},\cite{Aoki:2013ldr},~\cite{Bernardoni:2014fva},~\cite{Carrasco:2013zta},\cite{Aoki:2013ldr}; 
    {\it FLAG 13} estimates are determined by HPQCD 13 for $N_f=2+1+1$, HPQCD 12a, HPQCD 12b and FNAL-MILC 11 for $N_f=2+1$ and ETMC 13 for 
    $N_f=2$;
    (c) Refs.~\cite{Dowdall:2013tga},~\cite{Aoki:2013ldr},~\cite{Christ:2014uea},~\cite{Christ:2014uea},~\cite{Aoki:2014nga},~\cite{Na:2012kp},~\cite{Bazavov:2011aa},~\cite{Aoki:2013ldr},~\cite{Bernardoni:2014fva},~\cite{Carrasco:2013zta},~\cite{Aoki:2013ldr};  
    {\it FLAG 13} estimates are determined by HPQCD 13 for $N_f=2+1+1$, HPQCD 12a and FNAL-MILC 11 for $N_f=2+1$ and ETMC 13 for 
    $N_f=2$. Results for $f_{Bs}$ and $f_B$ from Ref.~\cite{Aoki:2014nga} display somewhat bigger errors 
    than the results shown above, so we have not included them in the plots.}
    \label{fig:compar_deconst}
\end{figure}

Using the ratio method we have obtained  non-perturbative results extrapolated to the continuum limit for the $b$-quark mass 
and its ratio to the charm and the strange quark mass.   
Moreover we have evaluated in the continuum limit the pseudoscalar $B$-decay constants, $f_{Bs}$, $f_B$ and their ratio 
as well as the (double) ratio of the latter with $f_{K}/f_{\pi}$. It is worth mentioning that the ratios between the 
SU(3) breaking ratios  $(f_{Bs}/f_B) / (f_K / f_{\pi})$ computed in this paper and the one of  
$(f_{Ds}/f_D) /  (f_K / f_{\pi})=1.003(14)$ 
determined in Ref.~\cite{Carrasco:2014poa}, are both perfectly compatible with unity within the errors, indicating, thus, 
an almost negligible dependence on the quark mass. 
Our results, Eqs.~(\ref{eq:mb}), (\ref{eq:mbmc}), (\ref{eq:mbms}) and (\ref{eq:fB1}) -- (\ref{eq:fB}), 
 are of high precision with well controlled systematic uncertainties.  
In Figs.~\ref{fig:compar_mass} and \ref{fig:compar_deconst} we compare our results with the ones obtained by other 
lattice collaborations. Each panel  
includes determinations of the relevant observable that are carried out in the continuum limit by using unquenched lattice simulations with $N_f = 2, 2+1$ and $2+1+1$ 
dynamical quarks.   
\clearpage

\noindent {\bf Acknowledgements}\\
We warmly thank our colleagues of the ETM Collaboration for fruitful discussions.
\noindent We acknowledge the CPU time provided by the PRACE Research Infrastructure under the projects
PRA027 ``QCD Simulations for Flavor Physics in the Standard Model and Beyond'' and PRA067 ``First Lattice
QCD study of B-physics with four flavors of dynamical quarks" on the BG/P and BG/Q systems at the J\"ulich
SuperComputing Center (Germany) and at CINECA (Italy), and by the agreement between INFN and CINECA
under the specific initiative INFN-LQCD123 on the Fermi BG/Q system at CINECA (Italy).

\end{document}